\documentclass{article}

\usepackage{PRIMEarxiv}

\usepackage[utf8]{inputenc} 
\usepackage[T1]{fontenc}    
\usepackage{hyperref}       
\usepackage{url}            
\usepackage{booktabs}       
\usepackage{amsfonts}       
\usepackage{nicefrac}       
\usepackage{microtype}      
\usepackage{lipsum}
\usepackage{fancyhdr}       
\usepackage{graphicx}       
\usepackage[dvipsnames]{xcolor}
\usepackage{tikz}
\usepackage{makecell}
\usepackage{subcaption}
\usepackage{multirow}
\usepackage{soul}
\usepackage{empheq}

\newcommand{\half}{\frac{1}{2}}

\title{{Fast and Optimal Incremental Parametric Procedure for the Densest Subgraph Problem: An Experimental Study}
\thanks{\textit{\underline{Citation}}: 
\textbf{Authors. Title. Pages.... DOI:000000/11111.}} 
}

\author{
  Dorit S. Hochbaum\\
  Department of Industrial Engineering and Operations Research\\
  University of California, Berkeley\\
  USA\\
  \texttt{dhochbaum@berkeley.edu} \\
  \And
  Ayleen Irribarra-Cortes\\
  School of Engineering\\
  Universidad de Concepci\'on\\
  Chile\\
  \texttt{airribarra@inf.udec.cl}\\
  \And
  Olivier Goldschmidt\\
  Riverside County Office of Education\\
  USA\\
  \texttt{goldoliv@gmail.com}
  \And
  Roberto As\'in-Ach\'a\\
  Department of Informatics\\
  Universidad T\'ecnica Federico Santa Mar\'ia\\
  Chile\\
  \texttt{roberto.asin@usm.cl}
}

\begin{document}
\title{Fast and Optimal Incremental Parametric Procedure for the Densest Subgraph Problem: An Experimental Study}

\maketitle

\begin{abstract}
The Densest Subgraph Problem (DSP) is widely used to identify community structures and patterns in networks such as bioinformatics and social networks.
While solvable in polynomial time, traditional exact algorithms face computational and scalability limitations, leading to the adoption of faster, but non-optimal, heuristic methods.
This work presents the first experimental study of the recently devised Incremental Parametric Cut (IPC) algorithm, which is an exact method for DSP and other ``monotone ratio problems". 
Our findings demonstrate that IPC not only overcomes the limitations of previous exact approaches but also substantially outperforms leading state-of-the-art heuristics in both speed and solution quality. 
IPC's performance is also evaluated here for other ``monotone ratio problems" related to conductance, Cheeger constant and normalized cut.
For these, our experimental study on large-scale instances demonstrate exceptional computational speed.
In particular, comparing IPC with the ``fully parametric cut" algorithm, which is the only other 
efficient known optimization algorithm for such problems, demonstrate the superior performance of IPC.
We provide here code and benchmarks, establishing IPC as a fast, scalable, and optimal solution framework for densest subgraph and related monotone ratio problems. 
\end{abstract}

\section{Introduction}
\label{sec:introduction}
The Densest Subgraph Problem (DSP) is fundamental in discovering salient structures in networks, with significant applications in domains that include social network analysis \cite{sozio2010community,kumar1999trawling}, bioinformatics \cite{fratkin2006motifcut}, and real-time story identification \cite{angel2012dense}. 
DSP has also been applied in data mining tasks, including the discovery of multiple dense subgraphs and density-constrained subgraphs \cite{balalau2015finding,tatti2015density}. 
It has further been used for fraud and spam detection in online networks \cite{beutel2013copycatch}, 
and for efficient large-scale graph analysis \cite{10.14778/2536336.2536342}.

Densest subgraph is utilized as the one-class machine learning classification problem, \cite{one-classML2001}. One-class machine learning is common in document and text classification, anomaly detection in time-series data, \cite{anomaly-detection2021} and outlier detection applications, \cite{outlier-detection2000lof}.
The DSP is to find, in a given graph, a subset of nodes, so that the ratio of the total weight of the edges connecting nodes in the subset, divided by the weight of nodes in the subset, is maximum. 
In the unweighted version, the goal is to maximize the ratio of the number of edges between nodes of the subset, divided by the size of the subset. Formally, DSP is to identify in a graph, $G=(V,E)$ with nonnegative edge weights and nonnegative node weights, a subset of vertices $S \subseteq V$, such that \( \frac{C(S, S)}{q(S)} \) is maximized, where \(C(S,S)\) is the sum of the weights of the edges with both endpoints in $S$, and $q(S)$ denotes the sum of the weights of the nodes of $S$. 

The densest subgraph has been studied for over five decades. The classical solution approach relies on cut-based algorithms, which guarantee optimal solutions, \cite{picard1982network,goldberg1984finding,gallo1989fast}. 
The link to cut problems originates in the ``linearization" approach for solving general ratio problems:
To solve a general ratio maximization problem \(\max_{x \in F} \frac{f(x)}{g(x)}\), {where $F$ is the set of feasible solutions,} it is sufficient to have an oracle providing a yes/no answer to the following \textbf{$\lambda$-question}:

\begin{center}
\emph{Is there a feasible solution $x \in F$ such that $\frac{f(x)}{g(x)} > \lambda$?}
\end{center}

Or, equivalently,
\begin{center}
\emph{Is there a feasible solution $x \in F$ such that $f(x) - \lambda g(x) > 0$?}
\end{center}
The answer to the $\lambda$-question, can be derived from the optimal solution to the \textbf{$\lambda$-problem}:
\[
\max_{x \in F} \left[ f(x) - \lambda g(x) \right].
\]

If the maximum value is greater than zero, a feasible solution with a ratio value strictly greater than $\lambda$ exists. Otherwise, the answer is no. Specifically, if the maximum value is strictly less than zero, then there is no feasible solution with a ratio value greater than or equal to $\lambda$. If the answer is zero, then the corresponding optimal solution for the $\lambda$-question has a ratio value of $\lambda$, which is the maximum ratio.

For the densest subgraph problem, Picard and Queyranne, \cite{picard1982network}, showed that the $\lambda$-problem is solved as a s-t max-flow min-cut problem on a respective graph.  Specifically the minimum $s,t$-cut, henceforth referred to as {\em min-cut}, provides a partition
of the nodes of the graph to a source set and a sink set, where the source set is the optimal solution to the $\lambda$-problem.  It was shown recently, in \cite{hochbaum2024incremental,hochbaum2023unified}, that for any ratio problem with ``monotone" formulation, the $\lambda$-problem is solved as a min-cut on a respective graph.  (For a definition of monotone formulations see Subsection \ref{subsec:mincut}.)
These monotone ratio problems include the DSP and several problems  related to the conductance ratio problem, called conductance*, defined below.

An obvious use of the $\lambda$-problem oracle is to embed it in a binary search on the value of $\lambda$. For the DSP, this entails making $O(\log n+ \log W + \log Q)$ calls to the oracle, with $n=|V|$, the number of nodes in the graph, and $W$ and $Q$ the largest edge weight and the largest node weight respectively.
Another common use of the $\lambda$-problem oracle is what we refer to as {\em incremental} process,
with early known uses in \cite{isbell1956attrition} and \cite{dinkelbach1967nonlinear}.  This process
is initiated with a feasible solution, setting its respective ratio value to $\lambda$. In each iteration the solution is either a strict improvement, in which case the next $\lambda$ 
value is set to be the ratio of this improved solution.  
Otherwise, the current 
solution is proved to be optimal.
In general, the incremental approach does not run in polynomial time since the number of successive parameter values explored, can assume, in the worst case, all possible values of the ratio, $O(WQ)$, using the notation above.

In a major breakthrough, Gallo et al.\ \cite{gallo1989fast} showed that for DSP, {\em all} the $\lambda$-problems can be solved with the {\em fully parametric cut} procedure in the complexity of a single min-cut.
This relied on the property of the $s,t$-graph associated with the $\lambda$-problem of DSP, in  which the source-adjacent capacities are monotone non-decreasing functions of the parameter $\lambda$, and the sink-adjacent capacities are monotone non-increasing functions of $\lambda$ (or vice versa), and all other capacities are independent of $\lambda$.
An $s,t$-graph with capacities that depend on a single parameter that has this property is called a {\em parametric flow graph}.
When increasing the value of $\lambda$ in a parametric flow graph the source sets of the respective min-cuts can only expand by adding nodes.  This {\em nestedness} property implies that although $\lambda$ can take an infinite number of values, there are at most $n$ different solution sets, for $n$ the number of nodes in the graph.
As the parameter increases, the (smallest) value of $\lambda$ where the solution set changes by adding at least one node to the source set, is called a \emph{breakpoint}.  Therefore there are at most $n$ breakpoints.
The {\em fully parametric cut} procedure identifies {\em all} the breakpoints and the respective source sets, and sink sets, of the cuts, for any parametric flow graph.
In \cite{hochbaum2009dynamic} it was shown that all breakpoints attained with the fully parametric cut form a {\em concave envelope} (discussed in Section \ref{sec:concave}) in which the leftmost breakpoint is the densest subgraph. 
Therefore this algorithm solves the densest subgraph problem in the complexity of min-cut.
Other variants of the parametric procedure discussed in \cite{gallo1989fast} are the {\em simple parametric cut} in which a list of monotone increasing parameter ($\lambda$) values are given as input, and the procedure solves for all these values, and the incremental procedure.
Both work in the complexity of a single min-cut, with an additional linear complexity for each parameter value in the sequence.

{The complexity of the parametric procedures of \cite{gallo1989fast} depended on the "continuity" property of the push-relabel max-flow algorithm \cite{goldberg1988new}:
once the min-cut problem is solved for one value of the parameter, that solution is used by the algorithm, without modifying labels, as a warm start for the updated graph capacities for the next value of the parameter, without affecting the overall complexity.} 
In \cite{hochbaum2008pseudoflow} and \cite{hochbaum1998pseudoflow} a new algorithm for max-flow min-cut was introduced, the Hochbaum PseudoFlow (HPF) algorithm. It was shown that HPF has the continuity property and the fully and simple parametric procedures, with HPF, run in the complexity of one min-cut procedure.
Note that the theoretical complexities of push-relabel and HPF are equivalent, and are denoted by $T(n,m)$ for a graph on $n$ nodes and $m$ arcs.

Despite the efficiency of the parametric cut procedures none of them have ever been used to solve DSP or other ``monotone" ratio problems enumerated in \cite{hochbaum2024incremental}. Instead, the dominant method has been the binary search such as the one proposed in \cite{goldberg1984finding}.
Consequently, a perception emerged of high computational cost of the polynomial time flow-based, or cut-based, methods \cite{boob2020flowless}.
With the increasing scale of DSP problems, 
research papers introduced heuristic methods 
that were to overcome the perceived high computational cost of the polynomial time cut-based methods.
Among early such heuristic algorithms is Charikar's greedy algorithm \cite{charikar2000greedy}, which provides a fast approximate solution. Recent advancements, including Greedy++ \cite{boob2020flowless} and FISTA~\cite{harb2022faster} algorithms, have improved upon Charikar’s approach, reporting to achieve a balance between efficiency and solution quality without requiring cut computations. 

Our study here focuses on the recently introduced  {\em Incremental Parametric Cut} (IPC), by Hochbaum \cite{hochbaum2024incremental,hochbaum2023unified}, for solving ``monotone ratio problems" that 
include DSP as a special case.  The algorithm utilizes  insights generated from the representation of the {\em concave envelope} of the breakpoints, \cite{hochbaum2009dynamic}, and in particular the fact that the leftmost breakpoint is the one providing the optimal ratio solution for any monotone ratio problem, whether maximization or minimization (for minimization, the envelope is convex).

Our study here also investigates the performance of the fully parametric cut.
{Prior to the work here, the fully parametric cut procedure has never been experimentally evaluated for ratio problems in general, and DSP in particular.
The only implementation available of the
fully parametric cut procedure is based on HPF and is available at \cite{webHPFSimple}. 

Unlike the fully parametric cut, IPC does not identify {\em all} breakpoints, but only a restricted subset of the breakpoints that is guaranteed to include the optimal ratio breakpoint--the leftmost breakpoint. IPC's theoretical complexity, using HPF, is that of one min-cut procedure, the same as that of the fully parametric cut procedure.  
However, as shown here, its performance in practice is dramatically faster.

To provide insights on why the practical running time of IPC is faster than that of the fully parametric cut procedure
we note that in both procedures
there is
a linear computational overhead, $O(n)$, for each resetting of the capacities for a new value of the parameter $\lambda$.   This overhead 
complexity, for exploring $K$ breakpoints, is $O(Kn)$, which is dominated by $T(n,m)$.  However, in practice, HPF runs in linear time, $O(m)$, and therefore this overhead can be very significant for graphs with large number of breakpoints.
In our experiments on a variety of datasets, for both DSP and the ratio problem related to conductance, conductance*, the number of breakpoints
explored by IPC is in the range of $2$ to $14$, which is often less than a fraction of a percent of the total number of breakpoints found by the fully parametric procedure. 
For some of the very large datasets, the fully parametric procedure's run time is in the order of magnitude of several hours, whereas IPC's run times is a few seconds, and below 209 seconds for all datasets.

In addition to DSP
we also provide a performance analysis of IPC for other ratio problems related to {\em minimum conductance} \cite{Jerrum-Sinclair-approx}, {\em normalized cut} \cite{shi2000normalized,Sis2003image}, {\em Cheeger constant}  \cite{cheeger1970lower} and to the {\em expansion ratio} \cite{alon1986eigenvalues} of a graph .
These problems are all NP-hard and there is a considerable body of literature on devising good quality approximations or heuristics for these problems.  
However, polynomial time solvable (monotone) ratio problems linked to these have been shown to run fast and deliver high quality solutions for some of these problems, e.g.\ \cite{hocNC2010}. To define the problems formally, we 
consider a graph, possibly directed, $G=(V,E)$, with nonnegative edge weights $w_{ij}$ and nonnegative node weights $q_i$.
The conductance problem is to find a subgraph $S\subset V$, subject to a ``size restriction" that minimizes the ratio of the cut capacity between $S$ and $\bar{S}$, denoted by $C(S,\bar{S})$, divided by
the weight of the nodes in $S$. The conductance problem is $\min _{ q(S)\leq \frac{q(V)}{2}} \frac{C(S,\bar{S})}{q(S)}$, and the expansion ratio problem is $\min _{|S|\leq \frac{|V|}{2}} \frac{C(S,\bar{S})}{|S|}$.  The special case in which the node weights are equal to the weighted degree of the node, $q_i=d_i$, the value  $\min _{ d(S)\leq \frac{d(V)}{2}} \frac{C(S,\bar{S})}{d(S)}$
is known as the {\em Cheeger constant}, \cite{cheeger1970lower}.
All these problems are NP-hard due to the ``size-restriction" constraint, $q(S)\leq \frac{q(V)}{2}$, or $|S|\leq \frac{|V|}{2}$.
Without the ``size-restriction" constraint, all these problems, as well as a number of others, are ``monotone" ratio problems, and as such were proved in \cite{hocNC2010,hoc-OR2013rayleigh}
to be polynomial-time solvable, and in particular with IPC, \cite{hochbaum2023unified,hochbaum2024incremental}.  

Conductance* is defined to be the conductance problem with the size restriction constraint relaxed.  However, simply omitting the size restriction results in the problem $\min _{ S\subseteq V} \frac{C(S,\bar{S})}{q(S)}$ which has a trivial  optimal solution, $S=V$ (since the cut value $C(V,\emptyset)$ is zero and the corresponding ratio value is $0$).  Without the size restriction the problem is of interest when the set $S$ is constrained to be a strict subset of $V$, say $V_1 \subset V$.   We then consider a set of ``seed" nodes $\bar{V_1} \subset V$ which is guaranteed {\em not} to be included in the set selected.  For such a selection of the seed nodes, conductance* is $\min _{ S\subseteq V_1} \frac{C(S,\bar{S})}{q(S)}$.
One way of using conductance* in order to approximate the solution to the conductance problem, is to find a subset of nodes $V_1$ which satisfies the size constraint, and is likely to contain the optimal solution.
There are various known practices for identifying a relevant set of seed nodes, and we will use one of those approaches in the experimental study.

Our experimental study for the conductance* problem demonstrates the efficiency of IPC for this minimization problem.  It also provides a stark comparison between the fully parametric cut procedure and 
IPC in terms of the number of breakpoints generated.
In a recent paper, \cite{conductance*2025}, we study methods for solving the NP-hard conductance, or Cheeger, problem, demonstrating that IPC, coupled with known methods to provide seed sets, delivers the best known solutions, and at faster running times, than all leading algorithms for these problems.

To summarize the contributions here, we
present a comprehensive experimental evaluation of the Incremental Parametric Cut (IPC) method, comparing its performance against leading heuristics for the Densest Subgraph Problem (DSP) and to the fully parametric cut procedure. The study applies to the DSP and 
the conductance* ratio problem, demonstrating IPC's superior performance across diverse graph sizes and structures. Through extensive testing on both weighted and unweighted graphs, we show that IPC outperforms recent heuristic approaches like Greedy++ and FISTA in speed, and obviously also in solution quality. Our experimental results are particularly notable for processing graphs with up to one-third of a billion edges in approximately three minutes, while providing, at termination, certified optimal solutions. This is in contrast to the other methods, such as Greedy++, that terminate after a prespecified number of iterations without information on the quality of the generated solution. This significant performance improvement positions IPC as a major advancement in solving the densest subgraph problem, with implications for various network structure analysis applications.

The remainder of this paper is structured as follows. Section~\ref{sec:prelims} introduces necessary notation and core concepts related to min-cut and parametric cut. In Section~\ref{sec:related_work} we discuss related work. 
Section~\ref{sec:ipc} reviews the IPC method for DSP and Conductance*. The results of our experimental study, along with implementation details, are presented in Section~\ref{sec:results}. In Section~\ref{sec:conclusions} we provide concluding remarks and suggest potential avenues for future research.

\section{Preliminaries}
\label{sec:prelims}
\subsection{Graph Definitions and Notations}

For an undirected graph $G=(V,E)$, with $V$ the set of nodes and $E$ the set of edges, each $[i,j]\in E$ with a positive weight $w_{ij}$, and two subsets of nodes $V_1,V_2 \subseteq V$,  we define the function $C(V_1,V_2)$ as $C(V_1,V_2) := \sum_{[i,j] \in E, i \in V_1, j \in V_2} w_{ij}$.  
In words, this is the sum of weights of edges that have one endpoint in $V_1$ and the other in $V_2$. For a directed graph $G=(V,A)$, where each arc $(i,j) \in A$ has a positive weight of $w_{ij}$, $C(V_1,V_2)$ is defined as $\sum_{(i,j) \in A, i \in V_1, j \in V_2} w_{ij}$, which is the sum of weights of the arcs that go from nodes in $V_1$ to those in $V_2$.
In particular $C(V_1, \bar{V_1})$ is the \textit{cut capacity} which is the sum of edge weights that have one endpoint in $V_1$ and the other in the complement $\bar{V_1}=V\setminus V_1$, or arcs' weights that go from $V_1$ to $\bar{V_1}$. 
Also, $C(V_1,V_1)$ is the sum of edge weights with both endpoints in $V_1$, or the sum of directed arc weights that have both their heads and tails in $V_1$.

Let the \textit{weighted degree} of node $i \in V$ be denoted by $d_i$ where $d_i = \sum_{[i,j] \in E, j \in V} w_{ij}$. The \textit{degree volume} of the set $S \subseteq V$ is denoted by $d(S)=\sum_{i \in S} d_i$.
For a directed graph, the \textit{weighted outdegree} \(d^+_i\) of a node \(i \in V\) is defined as \(d^+_i = \sum_{j|(i,j) \in A} w_{ij}\). For a set of nodes \(S \subset V\), we define its \textit{outdegree volume} of $S$ as \(d^+(S) = \sum_{i \in S} d^+_i\).
In either directed or undirected graph 
each node \(i \in V\) has an associated non-negative weight \(q_i\). 

\subsection{Min-Cut, Parametric Cut and monotone integer programs}
\label{subsec:mincut}
The min-cut problem is defined on a
directed $s,t$-graph \( G = (V\cup\{s,t\}, A) \), which features two designated nodes, $s$ and $t$, referred to as the source and sink nodes, respectively. Each arc is associated with a positive capacity value \( u_{ij} \). The min-cut problem is to partition $V$ into two subsets $S$ and $\bar{S}=V\setminus S$, called source set and sink set respectively, such that 
the sum of capacities going from $S\cup \{s\}$ to $\bar{S}\cup \{t\}$, $C(S\cup \{s\},\bar{S}\cup \{t\})$, is minimum. The min-cut problem is formulated as:

\begin{empheq}[box=\fbox]{align*}
    \min & & \sum_{(i,j)\in A} u_{ij}z_{ij} \\
    \text{s.t}  & & x_i - x_j \leq z_{ij} & & \forall (i,j) \in A \\ 
    & & x_s - x_t \geq 1  &  & \\
            & & z_{ij} \geq 0  &  & \forall (i,j) \in A \\
  \text{(min-cut)} & & x_t = 0  &  & 
\end{empheq}

There is an optimal solution $(x^*, z^*)$ that is binary (due to the total modularity of the constraint matrix) that partitions $V$ into two subsets: $S=\lbrace i \in V | x^*_i = 1\rbrace$ and $\bar{S}=\lbrace i \in V | x^*_i = 0\rbrace$.

It is noted that the min-cut problem is the dual of the maximum flow problem, max-flow.  The max-flow min-cut algorithm HPF solves firstly the min-cut problem, from which it can construct the maximum flow solution, if required. This maximum flow solution is not used in any of the applications described here and hence we only refer to the min-cut.

A {\em parametric flow graph} is an $s,t$-graph where the source adjacent arcs capacities are monotone non-increasing functions of a parameter $\lambda$, and the sink adjacent arcs capacities are monotone non-decreasing, or vice versa.

Given a parametric flow graph, the fully parametric min-cut procedure finds the min-cuts for all values of the parameter $\lambda$, in the complexity of a single min-cut. 

For a parametric flow graph with source adjacent arcs' capacities that are monotone non-increasing, and sink adjacent arcs' capacities that are monotone non-decreasing capacities, let $(S_{\lambda},\bar{S}_{\lambda})$ be the min-cut partition corresponding to the parameter value $\lambda$. For a sequence of parameter values $\lambda_0 < \lambda_1 < ... < \lambda_p$, and the corresponding min-cut partitions $(S_{\lambda_0},\bar{S}_{\lambda_0})$, $(S_{\lambda_1},\bar{S}_{\lambda_1})$, ..., $(S_{\lambda_p},\bar{S}_{\lambda_p})$, the source sets satisfy $S_{\lambda_0} \supseteq S_{\lambda_1} \supseteq ... \supseteq S_{\lambda_p}$, and the sink sets satisfy $\bar{S}_{\lambda_0} \subseteq \bar{S}_{\lambda_1} \subseteq ... \subseteq \bar{S}_{\lambda_p}$.  This \textit{nestedness} property was proved as a corollary of the parametric cut algorithms, in \cite{gallo1989fast,hochbaum1998pseudoflow,hochbaum2008pseudoflow}. 
A value of $\lambda$ where at least one node is removed from the source set and added to the sink set is called a \textit{breakpoint}. Consequently, there can be at most $n$ breakpoints.

Hochbaum, in ~\cite{HOCHBAUM2002291}, showed that any problem that can be formulated as \textit{monotone integer program} can be solved to optimality in polynomial time with a min-cut procedure on a respective graph. 
Monotone integer problems (Integer Program Monotone, IPM) are those formulated as integer programming with constraints that have at most two variables, $x$-variables, appearing with opposite sign coefficients, and possibly a third variable, $z$-variable, that can appear in one constraint only.  A typical constraint of an IPM formulation is of the form:  $a_ix_i-b_jx_j\leq c_{ij}+z_{ij}$, for $a_i$ and $b_i$ both nonnegative.  
A \textit{monotone integer program} (IPM) formulation for a set of $n$ $x$-variables and a set of constraints $A$, each associated with a pair of 
$x$-variables is as follows: 

\
\begin{empheq}[box=\fbox]{align*}
    \max & & \sum_{i=1}^{n} w_i x_i  - \sum_{(i,j) \in A} e_{ij}z_{ij}  \\
\text{s.t}  & & a_{ij}x_i -b_{ij}x_j \leq c_{ij} + z_{ij}& &  \forall (i,j) \in A \\ 
  & & \ell_i  \leq x_{i} \leq u_{i}, ~\text{integer} & & \forall i \in V \\ 
  \text{(IPM)} & & z_{ij} \geq 0, ~\text{integer} & & \forall (i,j) \in A  
\end{empheq}
All IPMs are solved as a min-cut on a $s,t$-graph derived from the formulation, which for binary problems has one node for each $x$-variable, and one arc for each constraint.
Furthermore, it was proved, in \cite{hocNC2010} and \cite{hoc-OR2013rayleigh}, that any ratio problem that has a monotone formulation with a respective parametric flow graph, is solved in the complexity of a single min-cut procedure using the fully parametric cut procedure. 

Note that constraints that do not include a $z$-variable, can still appear in this form with the respective coefficient of $z_{ij}$ in the objective function, $e_{ij}$, set to infinity.
The objective function coefficients of the $z_{ij}$ variables, $e_{ij}$ here, must be nonnegative for maximization and non-positive for minimization.

Any IPM problem was proved to be equivalent to a binary \textit{s-excess} problem, defined such that the variable $x_i = 1$ iff node $i$ is in the optimal set $S$. \\

\begin{empheq}[box=\fbox]{align*}
    \max & & \sum_{i\in V} w_ix_i - \sum_{(i,j)\in A} u_{ij}z_{ij} \\
    \text{s.t}  & & x_i - x_j \leq z_{ij} & & \forall (i,j) \in A \\ 
  & & x_j \in \lbrace 0, 1 \rbrace  &  & \forall \in \lbrace 1, ..., n \rbrace \\  
  \text{(s-excess)}& & z_{ij} \in \lbrace 0, 1 \rbrace  &  & \forall (i,j) \in A 
\end{empheq}

Note that the formulation of the min-cut problem is a specific case of \textit{s-excess}, where $w_i=0$.  As for min-cut, the constraint matrix of s-excess is totally unimodular and hence the integrality requirement on the variables can be relaxed (or disregarded).

As proved in~\cite{HOCHBAUM2002291}, the associated $s,t$-graph $G_{st}$ of this s-excess problem is constructed as follows: Each binary variable has an associated node in $V$;  nodes $s$ and $t$ are added; there is an arc from $s$ to every positive weight node $i$ with capacity $u_{si} = w_i$; there is an arc from every negative weight node $j$ to $t$ with capacity $u_{jt} = -w_j$;  every constraint in $A$ with $x_i$ and $x_j$ has an arc in the graph with capacity $u_{ij}$, which becomes infinite for constraints that involve only two variables. 
It was proved in~\cite{HOCHBAUM2002291} that $S^*$ is a set of maximum $s$-excess capacity in $G$ iff $S^*$ is the source set of a min-cut in $G_{st}$.

\section{Related Work}
\label{sec:related_work}
A sketch of the most relevant work related to the topics presented here has been provided in Section~\ref{sec:introduction}. Here we elaborate on these and other works.

\subsection{Cut-based Algorithms for DSP and other ratio problems}

Picard and Queyranne~\cite{picard1982network} were the first authors to link the $\lambda$-question with the min-cut problem and propose to address (the unweighted) DSP using cut techniques by performing $n$ successive calls to a min-cut procedure. This early method can be considered as the predecessor of the Incremental Parametric Cut, IPC, of \cite{hochbaum2024incremental}. 

Goldberg, in \cite{goldberg1984finding} used the binary search technique, which for the unweighted problem, reduced the computational complexity requiring $\mathcal{O}(\log n + \log m)$ calls to the min-cut subroutine. However, the algorithm is not strongly polynomial for weighted graphs.
The paper also made explicit use of a ``compact" formulation, proposed earlier, which reduced the size of the graph on which the min-cut problems were solved.

A substantial improvement was proposed by Gallo et al.\ \cite{gallo1989fast} in 1989, who devised the parametric max-flow min-cut algorithm. This algorithm used the push-relabel method to solve the maximum density subgraph problem in strongly polynomial time, 
leading to the most efficient complexity known for the problem, that of solving a single min-cut procedure on the associated graph.

The fully parametric cut procedure presented in \cite{gallo1989fast} employed the push-relabel max-flow algorithm from \cite{goldberg1988new}.
An alternative fully parametric cut procedure utilizing the Hochbaum PseudoFlow (HPF) algorithm \cite{hochbaum2008pseudoflow} was proposed in \cite{hochbaum2008pseudoflow} and \cite{hochbaum1998pseudoflow}, achieving computational complexity equivalent to a single min-cut procedure. Such complexity is implementation-dependent, denoted by $T(n,m)$. For instance, on current implementations $T(n,m) = O(mn \log \frac{n^2}{m})$ for both push-relabel \cite{goldberg1988new} and HPF \cite{hoc-Orlin2013HPF-improve}.
Both algorithms, push-relabel and HPF, posses the ``continuation'' property that enables their use in fully parametric cut procedures while maintaining the complexity of solving a single min-cut problem.
Among max-flow min-cut algorithms, only push-relabel and HPF are known to exhibit this ``continuation'' property.

A variant of the parametric cut procedure, discussed in \cite{gallo1989fast,hochbaum2008pseudoflow}
and implemented using HPF as described in \cite{chandran2009computational}, 
is the \textit{simple parametric} cut procedure.
Instead of finding all breakpoints, as in the fully parametric cut procedure, the input to simple parametric cut, is either a sequence of values of $\lambda$, or a sequence of capacities of source adjacent and sink adjacent arcs, such that the capacities are monotone non-decreasing on one side and monotone non-increasing on the opposite side (or vice versa). The procedure computes all min-cuts for the entire sequence. Like the \textit{fully parametric} cut procedure, the \textit{simple parametric} also takes advantage of the \emph{continuation} property to compute all solutions associated with the given sequence in the complexity of a single min-cut. The theoretical complexity of the simple parametric procedure is $O(T(n,m))+O(Kn)$, where $K$ is the number of values used as $\lambda$, given a range and a resolution. An implementation of the simple parametric cut, using HPF algorithm is available at \url{https://riot.ieor.berkeley.edu/Applications/Pseudoflow/parametric.html}.

Recently, in \cite{hochbaum2023unified}, Hochbaum presented a novel technique for solving any Lagrangian relaxation of NP-hard problems formulated as monotone integer programming (IPM) with a budget constraint. There, the author shows that the concave/convex envelope (for maximization/minimization) related to the Lagrangian relaxation of the budget constraint can be constructed as the output of a (parametric) min-cut procedure on a corresponding graph. It is further demonstrated that the 
Lagrangian relaxation of the budget constraint is the same as the $\lambda$-problem for the respective ratio problem, and the optimal solution to the ratio problem is the {\em leftmost} breakpoint in the envelope.
For monotone ratio problems it was proved that the Incremental Parametric Cut (IPC) procedure generates sequentially a small subset of all the breakpoints, ultimately reaching the leftmost breakpoint in the envelope, in the complexity of a single min-cut procedure. In the recent \cite{hochbaum2024incremental}, Hochbaum provided proof of the validity of IPC and demonstrated how the IPC method solves DSP, and many other known ratio problems. It was argued there that although the method has the same theoretical complexity as the fully parametric method of \cite{gallo1989fast,hochbaum1998pseudoflow,hochbaum2008pseudoflow}, it significantly reduces the number of the parameter values to test (breakpoints), which implies a considerable reduction in runtime in practice. 
Here we back this claim and experimentally show that this is indeed the case. Furthermore, we show that IPC outperforms fast heuristics that appeared in recent years by delivering optimal solutions much faster than those heuristics.

Lang and Rao~\cite{lang2004flow} addressed experimentally the conductance* problem, or more precisely, the Cheeger* and Expansion* Ratio problems.   To get a heuristic solution
that satisfies the size restriction, they employed \textsc{METIS}~\cite{karypis1998fast}—an approximation heuristic—to generate a seed set corresponding to one side of a balanced partition.
They showed that the $\lambda$-problem defined with the seed set fixed to be excluded from the solution set is solvable as a min-cut problem on a related graph.
Although this graph structure follows from the general theory on monotone integer programs, \cite{HOCHBAUM2002291}, they proved this result ad-hoc for the specific case of Cheeger*. (Note that although they refer to the conductance problem, they only address the case of node weights equal to weighted degree, $d_i$, or equal to $1$.)
Their method, called the ``Max Flow Quotient Cut Improvement'' (\textsc{MQI}), is closely related to IPC
as discussed in~\cite{hochbaum2024incremental}.
However, they did not prove, or claimed, that MQI finds the optimal solution and instead offered the alternative of using binary search in order to guarantee an optimal solution.
Also, their \textsc{MQI} approach required resolving a min-cut problem at each iteration, resulting in multiple calls to the procedure.
The authors showed experimentally that \textsc{MQI} outperformed the spectral method~\cite{hagen1992new} in computational speed while maintaining competitive solution quality. This implies that if MQI were to be replaced by IPC, the resulting procedure would be even much faster, by a significant factor, than the spectral method.

\subsection{Heuristics {and alternative algorithms} for the DSP}
Even though DSP has been known to be polynomial time solvable since the early 80s, there have been statements made of excessive computational time 
of the flow-based methods. This has motivated the introduction of heuristics for the problem.  Charikar \cite{charikar2000greedy} proposed a greedy algorithm that iteratively removes the vertex with minimum degree, proving this is a 2-approximation algorithm. The practical effectiveness and algorithmic simplicity have established this algorithm as a popular choice, particularly for applications requiring rapid approximations in dynamic or streaming graph environments \cite{mcgregor2015densest}.

Building upon the greedy framework, recent developments have introduced iterative refinements designed to improve solution quality while maintaining computational efficiency. The Greedy++ algorithm \cite{boob2020flowless} exemplifies this approach through the incorporation of iterative peeling layers that enhance the basic greedy mechanism. This enhancement yields superior approximations, particularly for graphs exhibiting skewed degree distributions. However, Greedy++ is known to produce suboptimal results for certain graph topologies \cite{tsourakakis2015k}.

A more recent contribution following this research direction is FISTA \cite{harb2022faster}, which introduces a novel iterative algorithm for the Densest Subgraph problem with a specialized implementation for unweighted graphs. The authors claim significant improvements over previous greedy algorithms, particularly Greedy++, in terms of computational efficiency and scalability for very large graphs. A key algorithmic innovation is the fractional peeling technique, which performs more effective rounding of fractional solutions compared to existing methods. The algorithm provides theoretical guarantees with provable $(1-\epsilon)$-approximations to optimal solutions and reportedly 
faster convergence than Greedy++. Greedy++ requires $O(\sqrt{m \Delta(G) / \epsilon})$ iterations, where $m$ denotes the number of edges and $\Delta(G)$ represents the maximum degree, with each iteration executable in $O(m)$ time.

While the empirical evaluation of FISTA on real and synthetic datasets claims substantial performance improvements over previous algorithms, our experimental results contradict these findings. This discrepancy arises from methodological differences in performance measurement. 
Specifically, the original paper excludes the time required for graph loading into memory and final solution computation from their reported execution times. More critically, their timing measurements do not account for the initialization overhead of data structures required for each algorithm iteration---only the core iteration time is tracked. In contrast, our evaluation methodology reports wall-clock time encompassing the complete computational process, from graph input through final solution output, ensuring a comprehensive and fair comparison across all methods.

Other works that addressed the detection of dense subgraphs include, \cite{liu2014dense-partition}, who propose the \emph{Dense Subgraph Partition} framework. This is an exact method that partitions a graph into an ordered list of so-called dense subgraphs. One component of this partition is the densest subgraph of the entire graph, which is identified at the termination of the algorithm.  With the reported run times, this is not a competitive method for finding densest subgraphs.  
We note that the Dense Subgraph Partition in \cite{liu2014dense-partition} consists of the difference sets between  the consecutive breakpoints of the concave envelope, which are each, optimal for its own given size (or budget). 
(In the paper they called these {\em critical $k$-sets}.)
These difference sets are {\em not} necessarily dense, but rather they are dense jointly with their predecessor breakpoints.
(We were unable to access the software for this algorithm.)

In an earlier work, the same authors introduced a heuristic algorithm called \emph{Shrinking and Expansion Algorithm} (SEA)~\cite{liu2013denseSEA} to generate a partition into dense subgraphs. 
The software for SEA is available and we ran it on several datasets to compare to the performance of IPC.
For all datasets tested, SEA reported a densest subgraph it found, which even for given seed nodes from the true densest subgraph, has considerably worse density than the true maximum density.
Also, the running times in all cases were hundreds of times more than the respective runtimes of IPC.

\section{Incremental Parametric Cut for DSP and Conductance*}
\label{sec:ipc}
{In this section we provide the monotone formulations of DSP and Conductance* as well as the description of how the IPC method applies to both problems.}

\subsection{The Densest Subgraph Problem, a {monotone formulation}}
\label{sec:DSP}
The Densest Subgraph Problem (DSP) is defined on an undirected graph $G=(V,E)$ with edge weights and node weights.  The goal is to find a subset of nodes \(S \subseteq V\) that maximizes the density of the subgraph induced by \(S\). The density 
is the ratio of the sum of the edge weights within \(S\) to the sum of the node weights in \(S\), formally,  \( \max _{S\subseteq V} \frac{C(S, S)}{q(S)} \). In the special case of unweighted graphs, which is the case most commonly studied in the literature, the edge weights and node weights are equal to $1$. 
We note that the problem is meaningful on directed graphs as well.

The $\lambda$-problem associated with DSP is
\[
\max_{S \subseteq V} \left[ C(S,S) - \lambda~q(S) \right].
\]
It was proved in \cite{hochbaum2024incremental} that this $\lambda$-problem for DSP is IPM and has a ``compact" monotone formulation that leads to a respective graph on $n+2$ nodes and $m+2n$ arcs as compared to previously known smallest graph formulation 
that required $n+2$ nodes and $2m+2n$ arcs.  Since $m=|E|$, the number of edges in the graph, is in general much larger than $n=|V|$, the number of nodes in the graph, this compact formulation affects the computational performance.  In particular, HPF runs, in practice, in linear time in the number of arcs in the graph, and therefore this compact formulation tends to improve the run times by a factor of $2$ as compared to using the previously known smallest graph formulation.

The following monotone formulation was shown to be a valid formulation for the $\lambda$-problem \cite{hochbaum2024incremental} for DSP, where the undirected graph $G=(V,E)$ is replaced by a directed graph $G=(V,A)$ that has one arc $(i,j)$ for each edge $[i,j]\in E$ such that $i<j$.  That is, the arcs are directed from the lower index endpoint to the higher index endpoint.  Note that the direction could have been set to be the opposite, from higher index to lower index, without affecting the validity.
 
\begin{empheq}[box=\fbox]{align*}
\max        & & \mathclap{\sum_{i\in V} (d_i^+ -\lambda q_i) x_i -\sum _{(i,j)\in A} u_{ij} z_{ij}}  \\
\text{s.t.} & & x_{ i} -x_j  \leq z_{ij}   & &{\forall (i,j)\in A}  \\
            & &  x_j  \in \lbrace0,1\rbrace & &~\forall j\in V\\
\text{(DSP-$\lambda$-problem)}& & z_{ij} \in \lbrace0,1\rbrace & &~\forall  (i,j)\in A.
\end{empheq}

The associated parametric flow graph for this monotone $\lambda$-problem, given in~\cite{hochbaum2024incremental}, is depicted in Figure~\ref{fig:network_dsp}.
In this graph there is an arc $(i,j)$ with capacity $w_{ij}$ for each edge $[i,j]$ in $E$ with $i<j$, and each node \(i\) has two arcs: one from the source $s$ and one to the sink $t$, with the following capacities:
\begin{itemize}
    \item From \(s\) to \(i\): an arc with capacity \(\max \{ 0, d_i^+ - \lambda q_i\}\) 
    \item From \(i\) to \(t\): an arc with capacity \(\max \{ 0, \lambda q_i - d_i^+ \}\) 
\end{itemize}

\begin{figure}[h]
    \centering
    \resizebox{.7\linewidth}{!}{
    \begin{tikzpicture}
\tikzset{>=latex}	
		\node [draw,circle] (s) at (0.0, 0.0) {s};
		\node [draw,circle] (t) at (8.0, 0.0) {t};

		\node [draw,circle] (1) at (4, 2.0) {1};
		\node [draw,circle] (n) at (4, -2.0) {n};
  
		\node [draw,circle] (i) at (4, 0.75) {i};
		\node [draw,circle] (j) at (4, -0.75) {j};
        \draw[->] (s) edge node[font=\tiny, above=-0.25em,sloped] {$max \{0, d_1^+ - \lambda q_1 \}$}  (1.west);
        \draw[->] (s) edge node[font=\tiny, above=-0.25em,sloped] {$max \{0, d_i^+ - \lambda q_i \}$} (i.west);
        \draw[->] (s) edge node[font=\tiny, above=-0.25em,sloped] {$max \{0, d_j^+ - \lambda q_j \}$} (j.west);
        \draw[->] (s) edge node[font=\tiny, above=-0.25em,sloped] {$max \{0, d_n^+ - \lambda q_n \}$} (n.west);

        \draw[<-] (t) edge node[font=\tiny, above=-.25em,sloped] {$max \{0, \lambda q_1 - d_1^+\}$} (1.east);
        \draw[<-] (t) edge node[font=\tiny, above=-.25em,sloped] {$max \{0, \lambda q_i - d_i^+\}$} (i.east);
        \draw[<-] (t) edge node[font=\tiny, above=-.25em,sloped] {$max \{0, \lambda q_j - d_j^+\}$} (j.east);
        \draw[<-] (t) edge node[font=\tiny, above=-.25em,sloped] {$max \{0, \lambda q_n - d_n^+\}$} (n.east);

        \path (1) -- node[auto=false,sloped]{\ldots} (i);
        \path (i) -- node[auto=false,sloped]{\ldots} (j);
        \path (j) -- node[auto=false,sloped]{\ldots} (n);
        \draw[->] (i) edge[out=-45,in=45,->] node[font=\tiny, right] {$w_{i,j}$}  (j);
\end{tikzpicture}
    }
    \caption{The parametric flow graph used to solve the $\lambda$-problem for DSP, densest subgraph. 
  }
    \label{fig:network_dsp}
\end{figure}
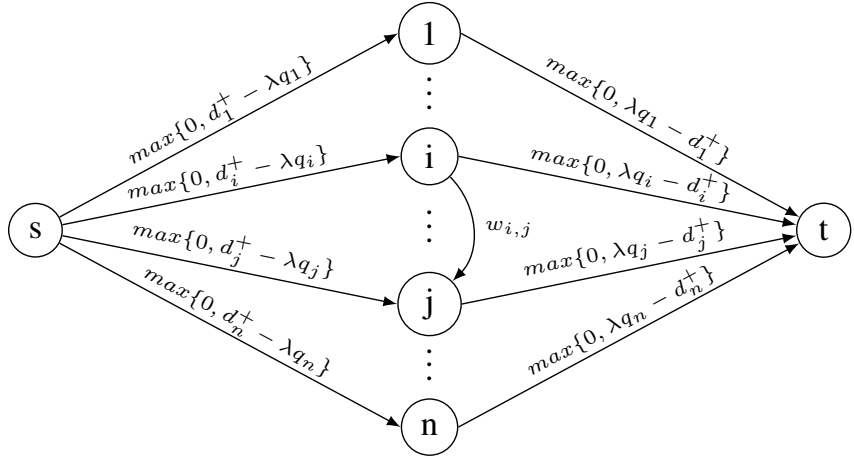

Since this is a parametric flow graph, it is possible to solve the respective DSP by employing the fully parametric cut procedure identifying the solutions for all possible values of $\lambda$. The solution set for DSP is given by the source set of the min-cut. Note that, since IPC is defined for parametric flow graphs having source adjacent arcs with non-decreasing capacities, the procedure is run on the reversed graph.

\subsection{Cheeger*, Expansion Ratio* and Conductance*  {monotone formulations}}
\label{sec:conductance}

Many ratio problems, defined for graphs with weighted nodes and weighted arcs, appear in contexts where the size of the optimal set is constrained, as discussed in Section~\ref{sec:introduction}.  
For an undirected graph $G=(V,E)$ with positive edge weights and positive node weights, the expansion ratio is defined as  \( \min _{S\subseteq V, |S|\leq \half |V|} \frac{C(S,\bar{S})}{|S|} \), and Cheeger's problem is defined as \( \min _{S\subseteq V} \frac{C(S,\bar{S})}{\min \{d(S),d(\bar{S}\}} \) and its optimal value is often referred to as the Cheeger constant.
Cheeger's problem is equivalent to \( \min _{S\subseteq V, d(S)\leq \half d(V)} \frac{C(S,\bar{S})}{d(S)} \).
The conductance problem is \( \min _{S\subseteq V, q(S)\leq \half q(V)} \frac{C(S,\bar{S})}{q(S)} \).
Both expansion ratio and Cheeger's are special cases of conductance, where for expansion ratio $q_i =1$ and for Cheeger's $q_i =d_i$, for all $i\in V$.  All these three problems are NP-hard. 
Relaxing the size constraint renders these problems monotone ratio problems, or IPM, that are solvable in polynomial time with a fully parametric cut procedure, \cite{hoc-OR2013rayleigh,hocNC2010}.
We name the size-relaxed problems {\em expansion ratio*} and {\em Cheeger*}.
Note that we address these problems for undirected graphs since all datasets considered here are undirected.  Yet the methods' extension to directed graphs is straightforward.

We now present the formulation of \textit{conductance*} as a parametric cut problem,  which applies also to both \textit{expansion ratio*} and \textit{Cheeger*}. 
The formulation utilizes a seed subset $\bar{V_0} \subset V$, so that its complement $V_0$ is guaranteed to contain the optimal solution. $V_0$ is selected so as to contain half the number of nodes.

The $\lambda$-problem for conductance* is:

\[
\min_{S \subseteq V_0} \left[ C(S,\bar{S}) - \lambda~q(S) \right].
\]

The parametric flow graph for conductance*'s $\lambda$-problem is derived from the monotone formulation of the problem, as shown in \cite{hochbaum2024incremental}, and is given in Figure~\ref{fig:network_conductance}.
This graph is generated as follows:  Given the seed set $\bar{V_0}$, each node $j \in \bar{V_0}$ is shrunk into the sink node.
Each node $i \in V_0$ is connected to the source node $s$ and sink node $t$ with arcs with capacities:

\begin{itemize}
    \item From $s$ to $i$: an arc with capacity $\lambda q_i$ 
    \item From $i$ to $t$: an arc with capacity $\sum_{k \in \bar{V_0}} w_{ik}$   
\end{itemize}

For each edge $[i, j]\in E$ with $i, j \in V_0$, there are two arcs $(i,j)$ and $(j,i)$ in the graph, both with capacity $w_{ij}$. 

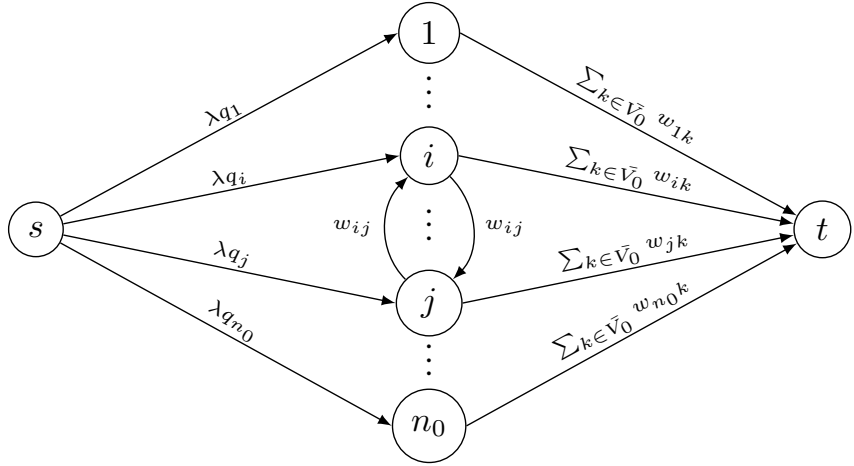
\begin{figure}[h]
    \centering
    \resizebox{.7\linewidth}{!}{
    \begin{tikzpicture}
\tikzset{>=latex}	
		\node [draw,circle] (s) at (0.0, 0.0) {$s$};
		\node [draw,circle] (t) at (8.0, 0.0) {$t$};

		\node [draw,circle] (1) at (4, 2.0) {$1$};
		\node [draw,circle] (n) at (4, -2.0) {$n_0$};
  
		\node [draw,circle] (i) at (4, 0.75) {$i$};
		\node [draw,circle] (j) at (4, -0.75) {$j$};
        \draw[->] (s) edge node[font=\tiny, above=-0.25em,sloped] {$\lambda q_1$}  (1.west);
        \draw[->] (s) edge node[font=\tiny, above=-0.25em,sloped] {$\lambda q_i$} (i.west);
        \draw[->] (s) edge node[font=\tiny, above=-0.25em,sloped] {$\lambda q_j$} (j.west);
        \draw[->] (s) edge node[font=\tiny, above=-0.25em,sloped] {$\lambda q_{n_0}$} (n.west);

        \draw[<-] (t) edge node[font=\tiny, above=-.25em,sloped] {$\sum_{k \in \bar{V_0}} w_{1k}$} (1.east);
        \draw[<-] (t) edge node[font=\tiny, above=-.25em,sloped] {$\sum_{k \in \bar{V_0}} w_{ik}$} (i.east);
        \draw[<-] (t) edge node[font=\tiny, above=-.25em,sloped] {$\sum_{k \in \bar{V_0}} w_{jk}$} (j.east);
        \draw[<-] (t) edge node[font=\tiny, above=-.25em,sloped] {$\sum_{k \in \bar{V_0}} w_{{n_0}k}$} (n.east);

        \path (1) -- node[auto=false,sloped]{\ldots} (i);
        \path (i) -- node[auto=false,sloped]{\ldots} (j);
        \path (j) -- node[auto=false,sloped]{\ldots} (n);
        \draw[->] (i) edge[out=-45,in=45,->] node[font=\tiny, right] {$w_{ij}$}  (j);
        \draw[->] (j) edge[out=135,in=-135,->] node[font=\tiny, left] {$w_{ij}$}  (i);
\end{tikzpicture}
    }
    \caption{The parametric flow graph for minimum conductance*'s $\lambda$-problem. Here $n_0=|V_0|$, the number of nodes in $V_0$.} 

    \label{fig:network_conductance}
\end{figure}

\subsection{Incremental Parametric Cut procedure and the Concave envelope} \label{sec:concave}

A general {\em monotone} ratio optimization problem on a feasible solution set $F$ is formulated as:
\[
\max_{x \in F} \frac{f(x)}{g(x)}.
\]
The numerator \(f(x)\) is referred to as the {\em benefit} and the denominator \(g(x)\), as the {\em budget}. Consider a function $f(x_B) = \max_{x \in F} \{f(x) \mid g(x) \leq B\}$ that associates a budget value $B$ with a maximum benefit. The \textit{concave envelope} is the lower envelope of all lines that lie above the set of optimal solutions, $(B,f(x_B))$ for any nonnegative $B$. 
A concave envelope is a piecewise linear function, which was proved in
\cite{hochbaum2009dynamic,hochbaum2023unified,hochbaum2024incremental} to have the following properties: The concave envelope and the breakpoints are found by the \textit{fully parametric} cut procedure; the breakpoints between the linear pieces are optimal solutions for their respective budgets; the solution sets of consecutive breakpoints are nested; there are most $n$ breakpoints;  and for ratio problems, the leftmost breakpoint is the optimal ratio solution.
An illustration of a concave envelope is given
in Figure~\ref{fig:concave_envelope}. 
\begin{figure}[h]
    \centering
    \begin{tikzpicture}[>=latex,thick]

\draw[->] (0,0) --  (6,0) node[below] {$B$};
\draw[->] (0,0) -- (0,3.2) node[left] {$f(x)$};

\node[draw,red,inner sep=2pt] (p0) at (0,0) {};
\node[draw,red,inner sep=2pt] (p1) at (2,1.98) {};
\node[draw,red,inner sep=2pt] (p2) at (2.5,2.31) {};
\node[draw,red,inner sep=2pt] (p3) at (3,2.57) {};
\node[draw,red,inner sep=2pt] (p4) at (3.5,2.77) {};
\node[draw,red,inner sep=2pt] (p5) at (5,3.04) {};

\node[fill=blue,circle,inner sep=2pt,scale=0.8] (p0) at (0,0) {};
\node[fill=blue,circle,inner sep=2pt,scale=0.8] (p1) at (2,1.98) {};
\node[fill=blue,circle,inner sep=2pt,scale=0.8] (p2) at (2.5,2.31) {};
\node[fill=blue,circle,inner sep=2pt,scale=0.8] (p3) at (3,2.57) {};
\node[fill=blue,circle,inner sep=2pt,scale=0.8] (p4) at (3.5,2.77) {};
\node[fill=blue,circle,inner sep=2pt,scale=0.8] (p5) at (5,3.04) {};

\node[fill=blue,circle,inner sep=2pt,scale=0.8] (b1) at (0.5,0.26) {};
\node[fill=blue,circle,inner sep=2pt,scale=0.8] (b2) at (1,0.46) {};
\node[fill=blue,circle,inner sep=2pt,scale=0.8] (b3) at (1.5,1.19) {};
\node[fill=blue,circle,inner sep=2pt,scale=0.8] (b4) at (4,2.77) {};
\node[fill=blue,circle,inner sep=2pt,scale=0.8] (b5) at (4.5,2.84) {};

\draw[red,thick] (p0) -- (p1) -- (p2) -- (p3) -- (p4) -- (p5);

\draw[black] (0,0) -- (p0);
\draw[black] (2,0) -- (p1);
\draw[black] (2.5,0) -- (p2);
\draw[black] (3,0) -- (p3);
\draw[black] (3.5,0) -- (p4);
\draw[black] (5,0) -- (p5);

\draw[black] (0.5,0) -- (b1);
\draw[black] (1,0) -- (b2);
\draw[black] (1.5,0) -- (b3);
\draw[black] (4,0) -- (b4);
\draw[black] (4.5,0) -- (b5);

\draw[->] (1.3,2.38) -- (1.3,1.98) -- (p1);
\node at (1.3,2.5) {\footnotesize $\frac{f(x^*)}{g(x^*)} = \max \frac{f(x)}{g(x)}$};

\begin{scope}[shift={(3.0,0.5)}] 
\draw[draw=black, inner sep=2ptm, fill=white] (-0.2,-0.2) rectangle ++(2.8,0.85);
  \node[draw,red,inner sep=2pt, right] (legendBox) at (0,0.35) {};
  \node[right] at (0.1,0.35) {\footnotesize Breakpoints};
  \node[fill=blue,circle,inner sep=2pt,scale=0.8] (legendDot) at (0.1,0.05) {};
\node[right] at (0.1,0) {\footnotesize Optimal solutions};
\end{scope}

\end{tikzpicture}
    \caption{The concave envelope, optimal solutions, breakpoints and ratio maximizing solution.}
    \label{fig:concave_envelope}
\end{figure}
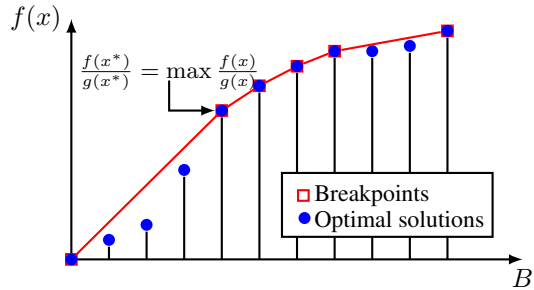

For a ratio {\em minimization} problem, $\min_{x \in F} \frac{f(x)}{g(x)}$, a \textit{convex envelope} is defined analogously as the upper envelope of all lines that bound the set of optimal solutions from below. Here, the leftmost \textit{breakpoint} of the \textit{convex envelope} minimizes the ratio $\frac{f(x)}{g(x)}$.  Examples of concave and convex envelopes for specific datasets are given in Figure \ref{fig:envelope-DSP} and Figure \ref{fig:envelope-conductance}, respectively. 

The Incremental Parametric Cut (IPC) is an algorithm
that computes an optimal solution for any monotone ratio problem.
Instead of identifying all breakpoints, as done with the fully parametric cut procedure, IPC finds a subsequence of breakpoints, ordered from right to left on the concave/convex envelope.
The number of breakpoints in this subsequence is typically a tiny fraction of the number of all breakpoints.
Like the simple and fully parametric cut procedures described in Section~\ref{subsec:mincut}, IPC's theoretical complexity is that of a single min-cut.  This is due exploiting the \emph{continuation} property of HPF, using the solution previously computed as a warm start for the updated capacities without modifying the labels. As shown here, because IPC explores a very small fraction of the total number of breakpoints, it performs in practice much faster. The comparison of the number of breakpoints and runtime, for DSP and conductance*, is reported in Table~\ref{tab:breakpoints} and Table~\ref{tab:breakpoints_conductance}. 
For example, for DSP's largest dataset, {\sc orkut}, IPC find the densest subgraph in less than four minutes as compared to seventy two hours for the fully parametric.

We now sketch the operation of IPC for a ratio maximization problem. Adapting it for ratio minimization problems is straightforward. 
IPC starts with a feasible solution set $S_0$ that is guaranteed to contain the optimal ratio solution set. In general, we set $S_0$ to be the entire graph.
For the value of $\lambda_0$, we can set it to the ratio (density) of the entire graph, or any other subgraph that has a different, possibly higher, ratio (density).
The default is to set $S_0=V$ and $\lambda_0 = \frac{f(S_0)}{g(S_0)}$, which is the ratio value for the entire graph. 

After setting up the initial value for the parameter, the $\lambda_0$-problem is solved, which either yields a solution $S_1\subset S_0$ with a strictly higher ratio, $\lambda_1$, or the associated value is zero, and thus is the optimal solution. The procedure continues iteratively  computing solution sets $S_0 \supset S_1 \supset ... \supset S_k $, 
and respective ratio values $\lambda_0 < \lambda_1 < ... < \lambda_k$, until no further improvement can be found. 

The key to the validity of IPC is the proof that it identifies, at each iteration, a {\em breakpoint} that is left to the current breakpoint solution, is strictly contained in it, and that the breakpoint solution set is guaranteed to contain the optimal solution set, \cite{hochbaum2024incremental}.

The pseudocode of IPC for DSP is given below. In the code {\bf HPF-para-continue}$(\lambda _k, S _k)$ refers to a subroutine of HPF that takes as input the source set $S_k$ previously computed, and its respective density $\lambda _k$ and {\em continue} to solve the $\lambda$-problem $\max  _{S\subseteq {S_k}} C(S,S) - \lambda _k q(S)$.
A general pseudocode for other ratio problems is given in \cite{hochbaum2024incremental}.

\vspace{0.1in}
\noindent
\begin{small}
{\sc Densest Subgraph Incremental parametric} ($S_0=V$, $\lambda _0$)
 \begin{description}
 \item[Step 0:]  Initialize $k=0$ 
\item[Step 1:] Use {\bf HPF-para-continue}$(\lambda _k, S _k)$ to solve,\\
$improve(\lambda _k)=\max  _{S\subseteq {S_k}} C(S,S) - \lambda _k q(S)$. \\
Let $S _{k+1}=\arg\max   _{S\subseteq {S_k}} C(S,S) - \lambda _k q(S)$.
\item[Step 2:] If $improve(\lambda _k)\leq 0$ stop. Output $S_{k}$. Else, continue
\item[Step 3:] \{$improve(\lambda_k)>0$\} Let $k := k+1$. 
\item[Step 4:] $\lambda _k =\frac{C(S_k,S_k) }{q(S_{k})}. $  Go to step 1.
\end{description}
\end{small}

In the case of \textit{conductance*}, $S_0$ is initialized as $V_0$, the complement of the seed set. 

\section{Implementation and experiments}
\label{sec:results}
We implemented IPC for DSP and conductance* in C. The code was compiled in gcc 14.2.1 with the optimization flag -O4. 
Our IPC code for DSP is available at~\cite{IPCforDSP}, and the IPC code for conductance* is available at~\cite{IPCforConductanceStar}. 
The Fully parametric HPF version used here is available at~\cite{WebHPFFull}. This implementation of the fully parametric procedure does not make use of all the features of the theoretical algorithm, e.g.\ {\em free runs}, and in that sense
does not match the theoretical complexity.  
However, as shown in this experimental study, the running time of the procedure is primarily dominated by the number of breakpoints it finds.

Since graph densities are rational numbers, which for our datasets have very large denominators, that can cause round-off errors.   In order to avoid this we introduce the option of choosing user-specified finite precision.
For the experiments here the precision is set up to four decimal places, i.e.\ with accuracy of $10^{-4}$.

\begin{table}[tt!]
\begin{subtable}[]{0.9\linewidth}
\centering
\begin{tabular}{lcrr}
\hline
\textbf{Dataset}            &\textbf{Source} & \textbf{Nodes} & \textbf{Edges} \\
\hline
\textsc{orkut}&\cite{konect}  & $ 8~730~857 $ & $ 327~036~486 $ \\
\textsc{trackers}&\cite{konect}  & $ 27~665~730 $ & $ 140~613~747 $ \\
\textsc{dbpedia-link}&\cite{konect}  & $ 18~268~991 $ & $ 126~890~209 $ \\
\textsc{wikidata}&\cite{hogan2019worst}  & $ 51~999~296 $ & $ 77~025~532 $ \\
\textsc{livejournal}&\cite{konect}  & $ 7~489~073 $ & $ 112~305~407 $ \\
\textsc{wiki-topcats}&\cite{snapnets}  & $ 1~791~489 $ & $ 25~444~207 $ \\
\textsc{cit-Patents}&\cite{snapnets}  & $ 3~774~768 $ & $ 16~518~947 $ \\
\textsc{actor-collab}&\cite{konect}  & $ 382~219 $ & $ 15~038~083 $ \\
\textsc{dblp-author}&\cite{snapnets}  & $ 4~000~150 $ & $ 8~649~002 $ \\
\textsc{ego-gplus}&\cite{snapnets}  & $ 107~614 $ & $ 12~238~285 $ \\
\textsc{web-BerkStan}&\cite{snapnets}  & $ 685~230 $ & $ 6~649~470 $ \\
\textsc{wiki-Talk}&\cite{snapnets}  & $ 2~394~385 $ & $ 4~659~565 $ \\
\textsc{flickr}&\cite{Zafarani+Liu:2009}  & $ 80~513 $ & $ 5~899~882 $ \\
\textsc{web-Google}&\cite{snapnets}  & $ 875~713 $ & $ 4~322~051 $ \\
\textsc{roadNet-CA}&\cite{snapnets}  & $ 1~965~206 $ & $ 2~766~607 $ \\
\textsc{com-youtube}&\cite{snapnets}  & $ 1~134~890 $ & $ 2~987~624 $ \\
\textsc{roadNet-TX}&\cite{snapnets}  & $ 1~379~917 $ & $ 1~921~660 $ \\
\textsc{roadNet-PA}&\cite{snapnets}  & $ 1~088~092 $ & $ 1~541~898 $ \\
\textsc{web-Stanford}&\cite{snapnets}  & $ 281~903 $ & $ 1~992~636 $ \\
\textsc{ego-twitter}&\cite{snapnets}  & $ 81~306 $ & $ 1~342~296 $ \\
\textsc{com-dblp}&\cite{snapnets}  & $ 317~080 $ & $ 1~049~866 $ \\
\textsc{com-amazon}&\cite{snapnets}  & $ 334~863 $ & $ 925~872 $ \\
\textsc{soc-Slashdot0902}&\cite{snapnets} & $ 82~168 $ & $ 504~230 $ \\
\textsc{soc-Slashdot0811}&\cite{snapnets} & $ 77~360 $ & $ 469~180 $ \\
\textsc{soc-Epinions1}&\cite{snapnets} & $ 75~879 $ & $ 405~740 $ \\
\textsc{email-Enron}&\cite{snapnets}  & $ 36~692 $ & $ 183~831 $ \\
\textsc{close-cliques}&\cite{harb2022faster} & $ 3~230 $ & $ 95 400 $ \\
\textsc{ego-facebook}&\cite{snapnets}  & $ 4~039 $ & $ 88~234 $ \\
\hline
\end{tabular}
\caption{Unweighted graphs}
\label{tables/unweighted-dataset}
\end{subtable}
\\
\vspace{2.5em}
\begin{subtable}[]{0.9\linewidth}
\centering
\begin{tabular}{lcrr}
\phantom{\textsc{soc-Slashdot0902}}&\phantom{\cite{konect}}  &  \phantom{$ 27~382~219$ }  & \phantom{$ 115~038~083 $} \\
\hline
\textbf{Dataset}            &\textbf{Source} & \textbf{Nodes} & \textbf{Edges} \\
\hline
\textsc{stackoverflow}&\cite{snapnets}  & $ 2~601~977 $ & $ 28~183~518 $ \\
\textsc{dblp\_coauthor}&\cite{konect}  & $ 1~824~701 $ & $ 8~344~615 $ \\
\textsc{twitter\_follow}&\cite{8919702}  & $ 2~248~044 $ & $ 5~932~351 $ \\
\textsc{twitter\_retweet}&\cite{8919702}  & $ 316~110 $ & $ 1~122~070 $ \\
\textsc{twitter\_reply}&\cite{8919702}  & $ 192~848 $ & $ 296~194 $ \\
\textsc{twitter\_quote}&\cite{8919702}  & $ 102~694 $ & $ 185~801 $ \\
\hline 
\end{tabular}
\caption{Edge-weighted graphs  }
\label{tables/weighted-dataset}
\end{subtable}\\[-0.5cm]
\caption{Datasets used in our experiments }
\label{tables/dataset}
\end{table}

For DSP, we compare the performance of IPC with Greedy++~\cite{boob2020flowless} and FISTA~\cite{harb2022faster}. Greedy++ is available at~\cite{GreedyPPCode} and FISTA is available in the supplemental material at ~\cite{FISTASupplement}. We only made minor modifications to enable fast input reading in FISTA and log wall clock time from the program's start to the end of each iteration. We note that the FISTA implementation does not report the running time for some parts of the code, such as data structure initialization and computing the density from their proxy solution at each iteration. Since these are important steps of the solution process and are part of the time reported by other algorithms, we also take them into account. We remark that the FISTA implementation is only designed for unweighted graphs; therefore, we did not run it for weighted graphs.

\begin{table}[ttt!]
\centering
\begin{tabular}{lrrrrr}
\hline
   & \multicolumn{1}{c}{\textbf{IPC}} & & & \multicolumn{1}{c}{\textbf{FISTA}}\\
   \multicolumn{1}{c}{\textbf{Dataset}} & \multicolumn{1}{c}{\textbf{optimal}} & \multicolumn{1}{c}{\textbf{IPC}} & \multicolumn{1}{c}{\textbf{FISTA}} & \multicolumn{1}{c}{\textbf{1st iter.}} \\
   & \multicolumn{1}{c}{\textbf{density}} & \multicolumn{1}{c}{\textbf{time}} & \multicolumn{1}{c}{\textbf{density}} & \multicolumn{1}{c}{\textbf{time}}\\ \hline
\textsc{orkut}               & $ 443.544 $                          & $ 208.981 $                       & 

$ (442.850) $                     & $ 922.916 $                           \\
\textsc{trackers}            & $ 334.477 $                          & $ 96.310 $                        & $ (332.704) $                     & $ 284.836 $                           \\
\textsc{dbpedia-link}        & $ 115.353 $                          & $ 100.753 $                       & $ (113.750) $                     & $ 245.385 $                           \\
\textsc{wikidata}            & $ 54.389 $                           & $ 86.946 $                        & $ (54.053) $                      & $ 149.156 $                           \\
\textsc{livejournal}         & $ 112.878 $                          & $ 68.738 $                        & $ (112.661) $                     & $ 213.732 $                           \\
\textsc{wiki-topcats}        & $ 71.931 $                           & $ 14.188 $                        & $ 71.931 $                      & $ 38.489 $                            \\
\textsc{cit-Patents}         & $ 40.129 $                           & $ 13.299 $                        & $ (39.080) $                      & $ 27.335 $                            \\
\textsc{actor-collab} & $ 309.303 $                          & $ 4.810 $                         & $ 309.303 $                     & $ 17.163 $                            \\
\textsc{dblp-author}         & $ 6.905 $                            & $ 10.837 $                        & $ (6.621) $                       & $ 14.994 $                            \\
\textsc{ego-gplus}           & $ 629.922 $                          & $ 3.246 $                         & $ 629.922 $                     & $ 13.777 $                            \\
\textsc{web-BerkStan}        & $ 103.406 $                          & $ 1.978 $                         & $ (100.495) $                     & $ 6.082 $                             \\
\textsc{wiki-Talk}           & $ 114.139 $                          & $ 2.919 $                         & $ 114.139 $                     & $ 10.389 $                            \\
\textsc{flickr}              & $ 444.752 $                          & $ 1.515 $                         & $ 444.752 $                     & $ 10.802 $                            \\
\textsc{web-Google}          & $ 28.041 $                           & $ 2.527 $                         & $ (27.787) $                      & $ 8.574 $                             \\
\textsc{roadNet-CA}          & $ 1.968 $                            & $ 3.272 $                         & $ (1.786) $                       & $ 7.265 $                             \\
\textsc{com-youtube}         & $ 45.599 $                           & $ 1.892 $                         & $ 45.599 $                      & $ 7.254 $                             \\
\textsc{roadNet-TX}          & $ 2.077 $                            & $ 2.432 $                         & $ (1.950) $                       & $ 4.258 $                             \\
\textsc{roadNet-PA}          & $ 1.878 $                            & $ 1.790 $                         & $ (1.698) $                       & $ 3.058 $                             \\
\textsc{web-Stanford}        & $ 59.390 $                           & $ 0.677 $                         & $ (58.493) $                      & $ 3.255 $                             \\
\textsc{ego-twitter}         & $ 69.622 $                           & $ 0.352 $                         & $ (69.410) $                      & $ 1.699 $                             \\
\textsc{com-dblp}            & $ 56.565 $                           & $ 0.561 $                         & $ 56.565 $                      & $ 1.611 $                             \\
\textsc{com-amazon}          & $ 4.804 $                            & $ 0.733 $                         & $ (4.308) $                       & $ 1.473 $                             \\
\textsc{soc-Slashdot}$ 0902 $     & $ 43.443 $                           & $ 0.154 $                         & $ 43.443 $                      & $ 0.570 $                             \\
\textsc{soc-Slashdot}$ 0811 $     & $ 42.271 $                           & $ 0.133 $                         & $ 42.271 $                      & $ 0.554 $                             \\
\textsc{soc-Epinions}$ 1 $        & $ 60.252 $                           & $ 0.117 $                         & $ 60.252 $                      & $ 0.455 $                             \\
\textsc{email-Enron}         & $ 37.344 $                           & $ 0.043 $                         & $ 37.344 $                      & $ 0.230 $                             \\
\textsc{close-cliques}       & $ 29.557 $                           & $ 0.010 $                         & $ (29.550) $                      & $ 0.086 $                             \\
\textsc{ego-facebook}        & $ 77.347 $                           & $ 0.009 $                         & $ 77.347 $                      & $ 0.133 $                            
\\ \hline
\end{tabular}
\caption{Running times in seconds and densities computed for IPC and one iteration of FISTA. Non-optimal densities found by FISTA are placed in parentheses. Only unweighted graphs are shown, since weights are not supported by FISTA.}
\label{table:first_iter}
\end{table}

\subsection{Experimental setup}
\label{ExperimentalSetup}

All the experiments were run on the Savio computational cluster at UC Berkeley.   Specifically, for each execution of each of the algorithms we evaluated, we requested a single node, consisting of a machine with Intel(R) Xeon(R) E5-2670 v3 CPU and 64 GB of main memory. The CPU has L3, L2, and L1 cache sizes of 30 MB, 256KB, and 32KB, respectively. Each machine runs Linux 4.18.0.  To assess the number of breakpoints with the fully parametric procedure, we use a machine with 256 GB of main memory.

\begin{table*}[hh!]
\small
\begin{subtable}[]{\linewidth}

\centering
\begin{tabular}{lrrrrrrrrrr}
\hline
                    & \multicolumn{1}{c}{\textbf{IPC}}            & \multicolumn{3}{c}{12 iterations of Greedy++}                  & \multicolumn{3}{c}{30 iterations of Greedy++}                  & \multicolumn{3}{c}{100 iterations of Greedy++}                 \\ \cmidrule(lr){2-2} \cmidrule(lr){3-5} \cmidrule(lr){6-8} \cmidrule(lr){9-11} 
\textbf{Dataset}             & \multicolumn{1}{c}{\textbf{Time [s]}} & \multicolumn{1}{c}{\textbf{Time [s]}} & \multicolumn{1}{c}{\textbf{Gap}} & \multicolumn{1}{c}{\textbf{SDF}} & \multicolumn{1}{c}{\textbf{Time [s]}} & \multicolumn{1}{c}{\textbf{Gap}} & \multicolumn{1}{c}{\textbf{SDF}}  & \multicolumn{1}{c}{\textbf{Time [s]}} & \multicolumn{1}{c}{\textbf{Gap}} & \multicolumn{1}{c}{\textbf{SDF}}  \\ \hline
\sc orkut               & $ 208.98 $                          & $ 679.08 $                   & $ 0.00\% $                   & $ 3.25\times $                    & $ 1600.88 $                  & $ 0.00\% $                   & $ 7.66\times $                    & $ 5200.73 $                  & $ 0.00\% $                   & $ 24.89 \times $                   \\
\sc trackers            & $ 96.31 $                           & $ 325.57 $                   & $ 0.00\% $                   & $ 3.38\times $                    & $ 753.37 $                   & $ 0.00\% $                   & $ 7.82\times $                    & $ 2402.23 $                  & $ 0.00\% $                   & $ 24.94 \times $                   \\
\sc dbpedia-link        & $ 100.75 $                          & $ 281.74 $                   & $ 0.00\% $                   & $ 2.80\times $                    & $ 662.85 $                   & $ 0.00\% $                   & $ 6.58\times $                    & $ 2146.94 $                  & $ 0.00\% $                   & $ 21.31 \times $                   \\
\sc wikidata            & $ 86.95 $                           & $ 192.99 $                   & $ 0.28\% $                   & $ 2.22\times $                    & $ 472.59 $                   & $ 0.10\% $                   & $ 5.44\times $                    & $ 1540.25 $                  & $ 0.00\% $                   & $ 17.72 \times $                   \\
\sc livejournal         & $ 68.74 $                           & $ 208.07 $                   & $ 0.00\% $                   & $ 3.03\times $                    & $ 463.37 $                   & $ 0.00\% $                   & $ 6.74\times $                    & $ 1457.54 $                  & $ 0.00\% $                   & $ 21.20 \times $                   \\
\sc wiki-topcats        & $ 14.19 $                           & $ 39.63 $                    & $ 0.00\% $                   & $ 2.79\times $                    & $ 97.68 $                    & $ 0.00\% $                   & $ 6.89\times $                    & $ 324.18 $                   & $ 0.00\% $                   & $ 22.85 \times $                   \\
\sc cit-Patents         & $ 13.30 $                           & $ 36.68 $                    & $ 0.00\% $                   & $ 2.76\times $                    & $ 90.63 $                    & $ 0.00\% $                   & $ 6.81\times $                    & $ 300.23 $                   & $ 0.00\% $                   & $ 22.57 \times $                   \\
\sc actor-collab& $ 4.81 $                            & $ 14.35 $                    & $ 0.00\% $                   & $ 2.98\times $                    & $ 35.26 $                    & $ 0.00\% $                   & $ 7.33 \times $                    & $ 117.67 $                   & $ 0.00\% $                   & $ 24.46 \times $                   \\
\sc dblp-author         & $ 10.84 $                           & $ 20.79 $                    & $ 1.47\% $                   & $ 1.92\times $                    & $ 50.56 $                    & $ 0.50\% $                   & $ 4.67 \times $                    & $ 166.48 $                   & $ 0.00\% $                   & $ 15.36\times  $                   \\
\sc ego-gplus           & $ 3.25 $                            & $ 13.89 $                    & $ 0.00\% $                   & $ 4.28\times $                    & $ 34.53 $                    & $ 0.00\% $                   & $ 10.64 \times $                   & $ 115.86 $                   & $ 0.00\% $                   & $ 35.70 \times $                   \\
\sc web-BerkStan        & $ 1.98 $                            & $ 4.18 $                     & $ 0.00\% $                   & $ 2.12\times $                    & $ 10.02 $                    & $ 0.00\% $                   & $ 5.07\times  $                    & $ 32.81 $                    & $ 0.00\% $                   & $ 16.59 \times $                   \\
\sc wiki-Talk           & $ 2.92 $                            & $ 6.78 $                     & $ 0.00\% $                   & $ 2.32\times $                    & $ 16.39 $                    & $ 0.00\% $                   & $ 5.62\times  $                    & $ 54.38 $                    & $ 0.00\% $                   & $ 18.63 \times $                   \\
\sc flickr              & $ 1.52 $                            & $ 5.23 $                     & $ 0.00\% $                   & $ 3.45\times $                    & $ 12.85 $                    & $ 0.00\% $                   & $ 8.48 \times $                    & $ 43.02 $                    & $ 0.00\% $                   & $ 28.39 \times $                   \\
\sc web-Google          & $ 2.53 $                            & $ 5.03 $                     & $ 0.00\% $                   & $ 1.99\times $                    & $ 12.09 $                    & $ 0.00\% $                   & $ 4.78 \times $                    & $ 39.36 $                    & $ 0.00\% $                   & $ 15.58 \times $                   \\
\sc roadNet-CA          & $ 3.27 $                            & $ 5.66 $                     & $ 0.67\% $                   & $ 1.73\times $                    & $ 14.46 $                    & $ 0.00\% $                   & $ 4.42 \times $                    & $ 48.64 $                    & $ 0.00\% $                   & $ 14.87 \times $                   \\
\sc com-youtube         & $ 1.89 $                            & $ 4.80 $                     & $ 0.01\% $                   & $ 2.54\times $                    & $ 11.66 $                    & $ 0.00\% $                   & $ 6.16 \times $                    & $ 38.41 $                    & $ 0.00\% $                   & $ 20.30 \times $                   \\
\sc roadNet-TX          & $ 2.43 $                            & $ 3.76 $                     & $ 0.00\% $                   & $ 1.55\times $                    & $ 9.39 $                     & $ 0.00\% $                   & $ 3.86 \times $                    & $ 31.61 $                    & $ 0.00\% $                   & $ 13.00 \times $                   \\
\sc roadNet-PA          & $ 1.79 $                            & $ 2.85 $                     & $ 2.62\% $                   & $ 1.59\times $                    & $ 7.05 $                     & $ 0.98\% $                   & $ 3.94 \times $                    & $ 23.70 $                    & $ 0.00\% $                   & $ 13.24 \times $                   \\
\sc web-Stanford        & $ 0.68 $                            & $ 1.55 $                     & $ 0.13\% $                   & $ 2.28\times $                    & $ 3.63 $                     & $ 0.06\% $                   & $ 5.37 \times $                    & $ 11.72 $                    & $ 0.00\% $                   & $ 17.32 \times $                   \\
\sc ego-twitter         & $ 0.35 $                            & $ 0.89 $                     & $ 0.00\% $                   & $ 2.52\times $                    & $ 1.88 $                     & $ 0.00\% $                   & $ 5.35 \times $                    & $ 5.97 $                     & $ 0.00\% $                   & $ 16.94 \times $                   \\
\sc com-dblp            & $ 0.56 $                            & $ 1.11 $                     & $ 0.00\% $                   & $ 1.98\times $                    & $ 2.63 $                     & $ 0.00\% $                   & $ 4.69 \times $                    & $ 8.58 $                     & $ 0.00\% $                   & $ 15.31 \times $                   \\
\sc com-amazon          & $ 0.73 $                            & $ 1.11 $                     & $ 2.10\% $                   & $ 1.52\times $                    & $ 2.66 $                     & $ 0.48\% $                   & $ 3.62 \times $                    & $ 8.63 $                     & $ 0.00\% $                   & $ 11.78 \times $                   \\
\sc soc-Slashdot$ 0902 $     & $ 0.15 $                            & $ 0.41 $                     & $ 0.00\% $                   & $ 2.64\times $                    & $ 0.90 $                     & $ 0.00\% $                   & $ 5.82 \times $                    & $ 2.85 $                     & $ 0.00\% $                   & $ 18.48 \times $                   \\
\sc soc-Slashdot$ 0811 $     & $ 0.13 $                            & $ 0.37 $                     & $ 0.00\% $                   & $ 2.79\times $                    & $ 0.83 $                     & $ 0.00\% $                   & $ 6.19 \times $                    & $ 2.63 $                     & $ 0.00\% $                   & $ 19.72 \times $                   \\
\sc soc-Epinions$ 1 $        & $ 0.12 $                            & $ 0.33 $                     & $ 0.00\% $                   & $ 2.81\times $                    & $ 0.72 $                     & $ 0.00\% $                   & $ 6.16 \times $                    & $ 2.29 $                     & $ 0.00\% $                   & $ 19.50 \times $                   \\
\sc email-Enron         & $ 0.04 $                            & $ 0.26 $                     & $ 0.00\% $                   & $ 6.14\times $                    & $ 0.44 $                     & $ 0.00\% $                   & $ 10.33 \times $                   & $ 1.15 $                     & $ 0.00\% $                   & $ 26.65 \times $                   \\
\sc close-cliques       & $ 0.01 $                            & $ 0.06 $                     & $ 0.02\% $                   & $ 6.25\times $                    & $ 0.09 $                     & $ 0.02\% $                   & $ 9.38 \times $                    & $ 0.16 $                     & $ 0.02\% $                   & $ 16.77 \times $                   \\
\sc ego-facebook        & $ 0.01 $                            & $ 0.09 $                     & $ 0.00\% $                   & $ 9.78\times $                    & $ 0.16 $                     & $ 0.00\% $                   & $ 17.17 \times $                   & $ 0.37 $                     & $ 0.00\% $                   & $ 40.00 \times $   \\

\hline 

\phantom{\sc twitter\_follow}  & \phantom{$ 3.227 $ }  & \phantom{$ 1162.220 $ }  &  & \phantom{$ 169.65 $ }  & \phantom{$ 1347.915 $ } &  & \phantom{$ 417.67 $ } & \phantom{$ 4461.008 $ } & \phantom{$ 0.00\% $ } & \phantom{$ 1382.32 $ } \\
\end{tabular}
\caption{Unweighted graphs}
\end{subtable}
\begin{subtable}[]{\linewidth}
\footnotesize
\centering
\begin{tabular}{lrrrrrrrrrr}
\phantom{\sc soc-Slashdot0902} & \phantom{$ 4.810 $ }                           & \phantom{$ 14.350 $ }                   & \phantom{$ 0.00\% $ }                  & \phantom{$ 2.98 $ }                   & \phantom{$ 35.260 $ }                   & \phantom{$ 0.00\% $ }                  & \phantom{$ 7.33 $ }                   & \phantom{$ 117.667 $ }                  & \phantom{$ 0.00\% $ }                  & \phantom{$ 24.46 $ }                  \\
\hline
                    & \multicolumn{1}{c}{\textbf{IPC}}            & \multicolumn{3}{c}{12 iterations of Greedy++}                  & \multicolumn{3}{c}{30 iterations of Greedy++}                  & \multicolumn{3}{c}{100 iterations of Greedy++}                 \\ \cmidrule(lr){2-2} \cmidrule(lr){3-5} \cmidrule(lr){6-8} \cmidrule(lr){9-11} 
\textbf{Dataset}             & \multicolumn{1}{c}{\textbf{Time [s]}} & \multicolumn{1}{c}{\textbf{Time [s]}} & \multicolumn{1}{c}{\textbf{Gap}} & \multicolumn{1}{c}{\textbf{SDF}} & \multicolumn{1}{c}{\textbf{Time [s]}} & \multicolumn{1}{c}{\textbf{Gap}} & \multicolumn{1}{c}{\textbf{SDF}}  & \multicolumn{1}{c}{\textbf{Time [s]}} & \multicolumn{1}{c}{\textbf{Gap}} & \multicolumn{1}{c}{\textbf{SDF}}  \\ \hline
 
\sc stackoverflow    & $ 16.86 $  & $ 1162.22 $  & $ 0.00\% $  & $ 68.95 \times $   & $ 2722.74 $  & $ 0.00\% $  & $ 161.53 \times $  & $ 8562.42 $  & $ 0.00\% $  & $ 507.98 \times $    \\
\sc dblp\_coauthor   & $ 5.05 $   & $ 413.80 $   & $ 0.00\% $  & $ 81.89 \times $   & $ 974.66 $   & $ 0.00\% $  & $ 192.89 \times $  & $ 3078.41 $  & $ 0.00\% $  & $ 609.22 \times $    \\
\sc twitter\_follow  & $ 3.28 $   & $ 547.48 $   & $ 0.00\% $  & $ 169.65 \times $  & $ 1347.92 $  & $ 0.00\% $  & $ 417.67 \times $  & $ 4461.01 $  & $ 0.00\% $  & $ 1382.32 \times $  \\
\sc twitter\_retweet & $ 0.48 $   & $ 17.70 $    & $ 0.00\% $  & $ 37.20 \times $   & $ 42.46 $    & $ 0.00\% $  & $ 89.25 \times $   & $ 132.55 $   & $ 0.00\% $  & $ 278.63 \times $    \\
\sc twitter\_reply   & $ 0.20 $   & $ 5.80 $     & $ 0.00\% $  & $ 28.68 \times $    & $ 13.83 $    & $ 0.00\% $  & $ 68.51 \times $   & $ 44.97 $    & $ 0.00\% $  & $ 222.74 \times $    \\
\sc twitter\_quote   & $ 0.11 $   & $ 2.02 $     & $ 0.00\% $  & $ 19.27 \times $   & $ 4.82 $     & $ 0.00\% $  & $ 45.99 \times $   & $ 15.62 $    & $ 0.00\% $  & $ 149.07 \times $                \\
\hline
\end{tabular}
\caption{Weighted graphs}
\end{subtable}
\caption{For instances of DSP, this table compares the performance of IPC and Greedy++ 
in terms of
running time in seconds, the gap of Greedy++ with the optimal solution delivered by IPC, and the slowdown factor (SDF) for 12, 30 and 100 iterations of Greedy++ respectively.}
\label{tab:complete}
\end{table*}

\subsection{Datasets}
\label{Datasets}

In our computational experiments, we employ a collection of all publicly available graphs previously used for DSP, and a few additional dataset. We include all graphs used in the experiments of~\cite{boob2020flowless}, except for two small graphs that are no longer available. 
These graphs are preprocessed in the same manner as in~\cite{boob2020flowless} in order to replicate their experiments. 
These graphs originate in the collections of the SNAP database~\cite{snapnets},  Koblenz Network Collection~\cite{konect}, and ASU's Social Computing Data Repository~\cite{Zafarani+Liu:2009}. 
Additional datasets used here include the \textsc{Close-Cliques} dataset from \cite{harb2022faster}; a set of Twitter graphs crawled from interactions during the first week of February 2018~\cite{8919702}; and a filtered dump of Wikidata used for benchmarking~\cite{hogan2019worst}.

The standard preprocessing techniques mentioned above, commonly used in the DSP literature, include: self-loops are removed, edge directions are disregarded, and multi-edges are replaced by a single weighted edge, where the weight of the edge is its multiplicity in the original multigraph. If the graph is reported as unweighted in the experiments, it means that all node and edge weights were set equal to $1$.
For all graphs in~\cite{boob2020flowless}, we followed the same criteria to classify them as weighted and unweighted graphs. As in ~\cite{harb2022faster}, we included {\sc close-cliques} as an unweighted graph. We added four new graphs that were not previously considered: ({\sc dbpedia-link}, {\sc wikidata}, {\sc stackoverflow} and {\sc dblp-coauthor}).  The first two are unweighted (node weights and edge weights equal to 1), and {\sc stackoverflow} and {\sc dblp-coauthor}) are edge weighted graphs.

For minimum conductance*, we chose only instances where the graphs are connected, since otherwise the optimal solution is trivial with a cut value of $0$. For a seed set, we use, as in~\cite{lang2004flow}, METIS~\cite{doi:10.1137/S1064827595287997}. METIS is used because it is a heuristic aimed at delivering a balanced cut partition, finding a close-to-minimum cut that partitions the graph into two almost-equal-sized partitions.  IPC is then run on the graph depicted in Figure~\ref{fig:network_conductance} where the output of METIS is the partition of $V$ to $V_0\cup \bar{V_0}$. Since in the datasets tested the nodes are unweighted, we use the node degrees $d_i$ in the input graph as node weights $q_i=d_i$.  Thus, effectively {solving} the Cheeger* problem.

We next report the computational results for both densest subgraph problem and the minimum conductance* problem. For the densest subgraph problem we compare the IPC performance results with those of FISTA and Greedy++, as well as to the fully parametric cut procedure. For the minimum conductance* problem we compare to the fully parametric cut and report running time results.

\subsection{The performance of IPC on the Densest Subgraph Problem}
We first present the experimental results for the Densest Subgraph Problem, which was defined on Section~\ref{sec:DSP}.

\begin{table*}[hhh!]

\begin{subtable}{\linewidth}
    
\centering
\begin{tabular}{lrrrrrr}
\hline

\textbf{Dataset}             & \multicolumn{1}{c}{\textbf{Breakpoints}} & \multicolumn{1}{l}{\textbf{Breakpoints}} & \multicolumn{1}{c}{\textbf{\% Explored}} & \multicolumn{1}{c}{\textbf{IPC}} & \multicolumn{1}{c}{\textbf{Fully Parametric}} & \multicolumn{1}{c}{\textbf{Runtime}} \\
                             & \multicolumn{1}{c}{\textbf{(Total \#)}} & \multicolumn{1}{c}{\textbf{(Explored \#)}} &  &\multicolumn{1}{c}{\textbf{Time [s]}} & \multicolumn{1}{c}{\textbf{Cut Time [s]}} & \multicolumn{1}{c}{\textbf{Fraction \%}}\\ \hline

\sc orkut            & 9347 & 8  & 0.09\%  & 208.981  & 260749.030 & 0.08\% \\
\sc trackers         & 5015 & 10 & 0.20\%  & 96.310   & 106574.925 & 0.09\% \\
\sc dbpedia-link     & 8606 & 12 & 0.14\%  & 100.753  & 166433.624 & 0.06\% \\
\sc wikidata         & 6231 & 12 & 0.19\%  & 86.946   & 168598.042 & 0.05\% \\
\sc livejournal      & 3309 & 8  & 0.24\%  & 68.738   & 35417.497  & 0.19\% \\
\sc wiki-topcats     & 3403 & 10 & 0.29\%  & 14.188   & 8438.242  & 0.17\% \\
\sc cit-Patents      & 6957 & 11 & 0.17\%  & 13.299   & 22020.681  & 0.06\% \\
\sc actor-collab     & 5396 & 8  & 0.15\%  & 4.810    & 4416.735  & 0.11\% \\
\sc dblp-author      & 1722 & 13 & 0.75\%  & 10.837   & 3988.703  & 0.27\% \\
\sc ego-gplus        & 2316 & 6  & 0.26\%  & 3.246    & 1206.426   & 0.27\% \\
\sc web-BerkStan     & 4140 & 9  & 0.22\%  & 1.978    & 1927.084   & 0.10\% \\
\sc wiki-Talk        & 656  & 9  & 1.37\%  & 2.919    & 712.444    & 0.41\% \\
\sc flickr           & 1593 & 6  & 0.38\%  & 1.515    & 404.403    & 0.37\% \\
\sc web-Google       & 3877 & 11 & 0.28\%  & 2.527    & 2230.664   & 0.11\% \\
\sc roadNet-CA       & 2711 & 12 & 0.44\%  & 3.272    & 2120.565   & 0.15\% \\
\sc com-youtube      & 919  & 9  & 0.98\%  & 1.892    & 520.333    & 0.36\% \\
\sc roadNet-TX       & 2154 & 12 & 0.56\%  & 2.432    & 1122.742   & 0.22\% \\
\sc roadNet-PA       & 1853 & 12 & 0.65\%  & 1.790    & 799.853   & 0.22\% \\
\sc web-Stanford     & 2194 & 8  & 0.36\%  & 0.677    & 405.677    & 0.17\% \\
\sc ego-twitter      & 1038 & 8  & 0.77\%  & 0.352    & 83.234    & 0.42\% \\
\sc com-dblp         & 1088 & 8  & 0.74\%  & 0.561    & 191.423    & 0.29\% \\
\sc com-amazon       & 1257 & 11 & 1.06\%  & 0.733    & 237.895    & 0.31\% \\
\sc soc-Slashdot0902 & 383  & 7  & 1.83\%  & 0.154    & 18.119     & 0.85\% \\
\sc soc-Slashdot0811 & 341  & 7  & 2.05\%  & 0.133    & 15.187     & 0.88\% \\
\sc soc-Epinions1    & 389  & 7  & 1.80\%  & 0.117    & 14.565     & 0.80\% \\
\sc email-Enron      & 358  & 7  & 1.96\%  & 0.043    & 4.212      & 1.02\% \\
\sc close-cliques    & 4    & 2  & 50.00\% & 0.010    & 0.068      & 14.71\% \\
\sc ego-facebook     & 196  & 5  & 2.55\%  & 0.009    & 0.517      & 1.74\% \\
 
\hline
\end{tabular}
\caption{Unweighted graphs}
\end{subtable}

\begin{subtable}{\linewidth}
    
\centering
\begin{tabular}{lrrrrrr}

\phantom{\textsc{soc-Slashdot0902}} & \phantom{2767} & \phantom{8}                   & \phantom{0.29\%}                  \\
\hline
                             \textbf{Dataset}             & \multicolumn{1}{c}{\textbf{Breakpoints}} & \multicolumn{1}{l}{\textbf{Breakpoints}} & \multicolumn{1}{c}{\textbf{\% Explored}} & \multicolumn{1}{c}{\textbf{IPC}} & \multicolumn{1}{c}{\textbf{Fully Parametric}} & \multicolumn{1}{c}{\textbf{Runtime}} \\
                             & \multicolumn{1}{c}{\textbf{(Total \#)}} & \multicolumn{1}{c}{\textbf{(Explored \#)}} &  &\multicolumn{1}{c}{\textbf{Time [s]}} & \multicolumn{1}{c}{\textbf{Cut Time [s]}} & \multicolumn{1}{c}{\textbf{Fraction \%}}\\ \hline

\sc stackoverflow    & 4449 & 13 & 0.29\% & 16.856 & 14226.817 & 0.12\% \\
\sc dblp\_coauthor   & 3900 & 9  & 0.23\% & 5.053  & 5348.319  & 0.09\% \\
\sc twitter\_follow  & 1010 & 9  & 0.89\% & 3.227  & 1137.485  & 0.28\% \\
\sc twitter\_retweet & 1180 & 8  & 0.68\% & 0.476  & 201.491   & 0.24\% \\
\sc twitter\_reply   & 524  & 9  & 1.73\% & 0.202  & 41.764    & 0.48\% \\
\sc twitter\_quote   & 343  & 8  & 2.33\% & 0.105  & 12.378    & 0.85\% \\
 
\hline
\end{tabular}
\caption{Edge-weighted graphs}
\end{subtable}

\caption{For DSP instances: The total number of breakpoints, delivered by fully parametric; the number of breakpoints explored by IPC; the percentage of breakpoints explored out of the total; and running times of fully parametric and of IPC for each dataset. }
    \label{tab:breakpoints}
\end{table*}

\subsubsection{Comparing the performance of IPC to that of FISTA and Greedy++}
Here we analyze and compare the practical performance of IPC for computing the densest subgraph against two recently proposed heuristic methods, respectively Greedy++~\cite{boob2020flowless} and FISTA~\cite{harb2022faster}.

First, we compare the solution quality and running times of IPC against one iteration of FISTA for computing the maximum density subgraph. We run FISTA for only one iteration since 
our experiments show that the runtime of IPC is always shorter than that of a single iteration of FISTA, as seen in Table~\ref{table:first_iter}.  
Since the available FISTA implementation only supports unweighted graphs our experiments here were restricted to such graphs. 
From the results in Table~\ref{table:first_iter} we conclude that since FISTA's single iteration is always slower than IPC and since it does not guarantee optimal solutions, 
it does not provide any advantage over IPC in any scenario. 

\begin{table*}[ttt!]
\footnotesize
\centering
\begin{tabular}{lrrccrrr}
\hline
\textbf{Dataset}             & \multicolumn{1}{c}{\textbf{Breakpoints}} & \multicolumn{1}{l}{\textbf{Breakpoints}} & \multicolumn{1}{c}{\textbf{Explored \%}}  & \multicolumn{1}{c}{\textbf{Minimum}} & \multicolumn{1}{c}{\textbf{IPC run}} & \multicolumn{1}{c}{\textbf{Fully Parametric}}  & \multicolumn{1}{c}{\textbf{Runtime}}\\
                             & \multicolumn{1}{c}{\textbf{(Total \#)}} & \multicolumn{1}{c}{\textbf{(Explored \#)}} &  \textbf{} & \multicolumn{1}{c}{\textbf{Conductance*}} & \multicolumn{1}{c}{\textbf{time [s]}} & \multicolumn{1}{c}{\textbf{Cut Time [s]}}& \multicolumn{1}{c}{\textbf{Fraction \%}}\\
                             \hline \hline
\sc livejournal$_M$      & 4036 & 12 & 0.30\%  & 0.0017 & 23.07 & 9784.301   & 0.24\% \\
\sc wiki-topcats$_M$     & 2    & 1  & 50.00\% & 0.0922 & 10.18 & 66.431     & 15.32\% \\
\sc ego-gplus$_M$        & 85   & 5  & 5.88\%  & 0.0797 & 1.69  & 19.248     & 8.78\% \\
\sc flickr$_M$           & 1224 & 4  & 0.33\%  & 0.1096 & 0.39  & 56.365     & 0.69\% \\
\sc com-youtube$_M$      & 398  & 8  & 2.01\%  & 0.0044 & 2.14  & 147.924    & 1.45\% \\
\sc ego-twitter$_M$      & 22   & 5  & 22.73\% & 0.0082 & 0.32  & 2.010      & 15.92\% \\
\sc com-dblp$_M$         & 197  & 8  & 4.06\%  & 0.0041 & 0.53  & 16.809     & 3.15\% \\
\sc com-amazon$_M$       & 302  & 6  & 1.99\%  & 0.0006 & 0.38  & 27.841     & 1.37\% \\
\sc soc-Slashdot0902$_M$ & 33   & 6  & 18.18\% & 0.0189 & 0.11  & 0.839      & 13.11\% \\
\sc soc-Slashdot0811$_M$ & 34   & 6  & 17.65\% & 0.0232 & 0.11  & 0.727      & 15.13\% \\
\sc ego-facebook$_M$     & 85   & 2  & 2.35\%  & 0.0013 & 0.01  & 0.049      & 20.41\% \\

\hline
\end{tabular}
\caption{For minimum conductance* instances:  The total number of breakpoints, delivered by fully parametric; the number of breakpoints explored by IPC; the percentage of breakpoints explored out of the total; the minimum conductance* values, delivered by IPC, and running times of fully parametric and of IPC for each dataset.}
    \label{tab:breakpoints_conductance}
\end{table*}

We next compare, in Table~\ref{tab:complete}, IPC and Greedy++. The number of iterations of Greedy++ is determined as per the 
authors' recommendation in~\cite{boob2020flowless}: It is  suggested to run Greedy++ for 12, 30, or 100 iterations, with the rationale that 12 is the average number of iterations that were required to compute the optimum on the datasets they used;
30 iterations were required, again, for the datasets they used, to compute a solution within $1\%$ of the optimum; and 100 iterations were required to attain the optimum on all datasets they used.   
Note that our results show that for one dataset that \cite{boob2020flowless} did not use, {\sc close-cliques}, even after 100 iterations, the maximum density subgraph was not found. This was already pointed out by~\cite{harb2022faster}.
In Table~\ref{tab:complete} we present the running times
and solution quality gaps (as compared with running time and the optimal value attained by IPC) for 12, 30, and 100 iterations of Greedy++.

By analyzing Table~\ref{tab:complete}, we again note that the running time of IPC is always faster than running Greedy++ for the recommended number of iterations while certifying the optimality of the solution. For unweighted graphs, we achieve a speedup by a factor of $2$ to $3$. A much higher factor is attained for weighted graphs, with one or two orders of magnitude speedup. This substantial difference between unweighted and weighted graphs can be attributed to the sophisticated data structures that Greedy++ employs specifically for weighted graph processing.

\subsubsection{Comparison of IPC with fully parametric procedure}

\begin{figure}[hhh!]
    \centering
    \includegraphics[width=0.5\linewidth]{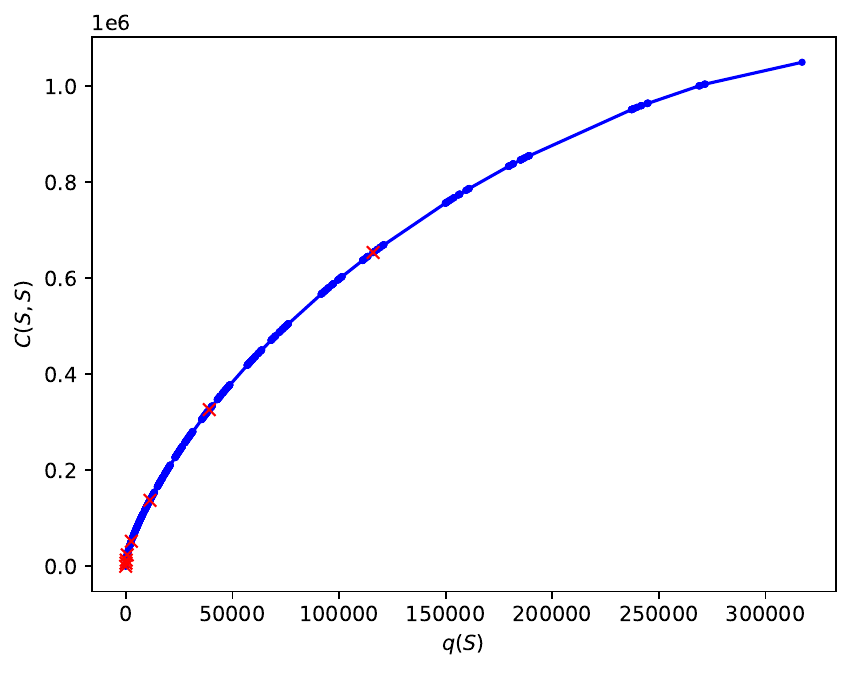}
    \label{fig:subfigA-DSP}\\[-0.25cm]
    \caption{All breakpoints in the concave envelope ({\color{blue}$\bullet$}) and the subset of breakpoints ({\color{red}$\times$}) explored by IPC, for DSP on \textsc{com-dblp} dataset.}
    \label{fig:envelope-DSP}
\end{figure}

We compare the performance of IPC against the fully parametric cut procedure. To this end, we measure the runtime of the fully parametric cut procedure and assess the number of breakpoints that IPC explores in contrast to the total number of breakpoints generated by the fully parametric procedure.

The interval for the values of the parameter $\lambda$ used as input for the fully parametric procedure lies between zero and half of the maximum weighted degree. This interval serves as a loose upper bound on the maximum density of a subgraph. It is worth noting that a narrower interval—such as one ending with the density of the entire graph—can also be used.

The results, presented in Table~\ref{tab:breakpoints}, demonstrate that the number of breakpoints explored by IPC is consistently small, never exceeding thirteen for all datasets tested. Moreover, this number represents only a very small fraction of the total number of breakpoints, which, in most cases, is one or two orders of magnitude larger.

This substantial disparity in the number of breakpoints explored by the fully parametric algorithm versus IPC translates into a significant difference in runtime, as observed on the graphs tested here. These graphs are, in general, among the largest studied in the literature. This runtime difference is evident from the results shown in Table~\ref{tab:breakpoints}. 
We also observe a strong correspondence between the fraction of breakpoints explored by IPC and its runtime relative to the fully parametric cut procedure. This indicates that, although the fully parametric implementation lacks features required to achieve its theoretical complexity, the primary factor driving the runtime difference is the significantly smaller subset of breakpoints explored by IPC.

Figure~\ref{fig:envelope-DSP} provides a compelling visualization of this efficiency contrast for the \textsc{com-dblp} dataset, showing the concave envelope with both the complete set of breakpoints and those selectively explored by IPC. The breakpoints examined by IPC (marked as red crosses) represent merely 0.74\% of the total breakpoints (shown as blue dots). This enables IPC to achieve remarkable computational efficiency, requiring only 0.21\% of the running time needed by the fully parametric cut procedure—a performance improvement exceeding two orders of magnitude.

\subsection{Performance of IPC for Conductance*}

In this section, we present our results on the minimum conductance* problem, defined in Section~\ref{sec:conductance}.  
Recall that conductance* problem has a trivial solution of value $0$ if the graph is not connected, or if there is no seed set of nodes that is to be excluded from the solution set.
For that purpose 
we ran METIS~\cite{doi:10.1137/S1064827595287997} on the datasets that are connected graphs to compute the subset on which conductance* is computed. The METIS' graph partition tool \texttt{(gpmetis)} was run  with the following parameters:  
{\ttfamily ptype=rb}, 
{\ttfamily objtype=cut}, 
{\ttfamily ctype=shem}, 
{\ttfamily rtype=fm}, 
{\ttfamily iptype=grow}, 
{\ttfamily dbglvl=0}, 
{\ttfamily ufactor=1.001}, 
{\ttfamily no2hop=NO}, \linebreak
{\ttfamily minconn=NO}, 
{\ttfamily contig=NO}, 
{\ttfamily ondisk=NO}, 
{\ttfamily nooutput=NO}, 
{\ttfamily seed=-1}, 
{\ttfamily niparts=-1}, 
{\ttfamily niter=10}, 
{\ttfamily ncuts=1}.
We append the subscript $M$ to the graph name to indicate that it is the modified version subgraph computed with METIS.
None of the edge-weighted graphs in our datasets are included in the IPC for conductance* runs since they are all disconnected. From the two sets of the partition generated by METIS, we choose as the seed set the one that has the largest number of nodes.

\begin{figure}[ttt!]
    \centering
    \includegraphics[width=0.5\linewidth]{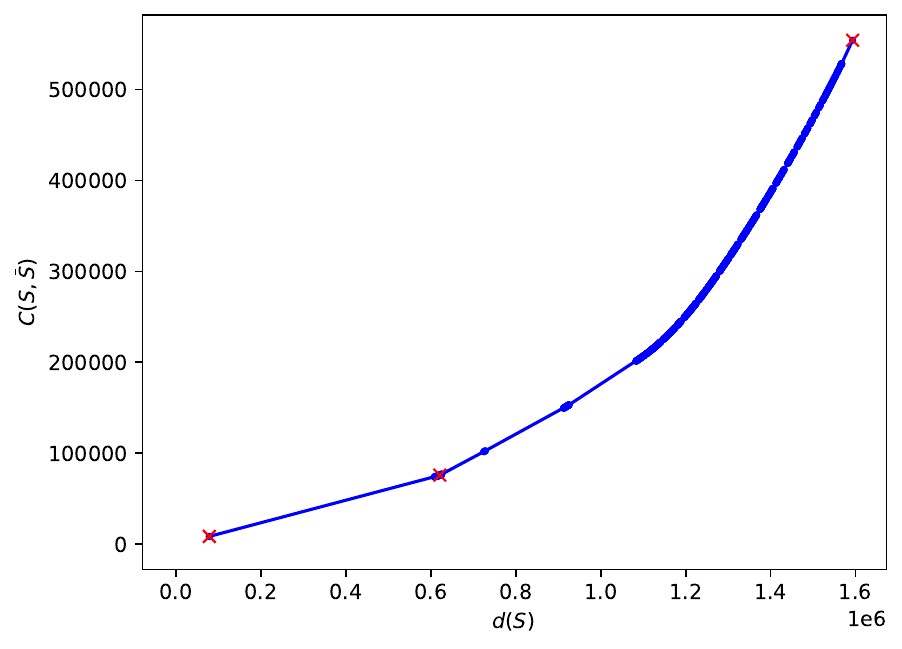}
	\label{fig:subfigA-cond}
	\caption{All the breakpoints of the convex envelope ({\color{blue}$\bullet$}) and the subset of breakpoints ({\color{red}$\times$}) explored by IPC, for conductance* on \textsc{flickr}$_M$ dataset.}
    \label{fig:envelope-conductance}
\end{figure}

We compare the performance of IPC for conductance* against the fully parametric cut procedure. We do so by measuring the runtime of the fully parametric cut procedure, as well as by assessing the total of number of breakpoints, as computed by the fully parametric procedure, and comparing with the number explored by IPC. 
In a related paper, \cite{conductance*2025}, we report on a direct comparison between IPC applied to conductance* and other leading algorithms for the conductance problem. As shown there, by performing multiple runs of the combination of METIS, creating seed sets, and IPC, we outperform, both in time and quality, leading methods applied to the conductance problem 
including the Spectral Method~\cite{cheeger1970lower} and LocalSpectral~\cite{fountoulakis2023flow}

In our experiments on the conductance* problem, the node weights are set as the weighted degrees, $q_i = d_i$. For these weights, the ratio $\frac{C(S, \bar{S})}{d(S)}$ for any subset $S \subseteq V$ is always less than $1$. 
Therefore, for the conductance* problem, we apply the fully parametric cut procedure to identify all breakpoints within the interval $[0, 1]$. It is worth noting that, for the purpose of solving the conductance* problem, the fully parametric procedure can be run on a narrower interval, $\left[ 0, \frac{C(V_0, \bar{V_0})}{q(V_0)} \right]$. The results are presented in Table~\ref{tab:breakpoints_conductance}.

For all graphs tested, IPC explores only a fraction of the lower bounds on the number of breakpoints, with the difference being more pronounced for graphs with a higher number of breakpoints. This results in a significant runtime advantage for IPC.
Here, we also observe that the fraction of breakpoints explored by IPC closely matches its share of the runtime relative to the fully parametric approach, just as observed for the densest subgraph problem. This reinforces our finding that the main source of IPC’s speedup is its ability to solve the problem optimally while considering only a small subset of breakpoints. Once again, although the fully parametric implementation does not achieve its theoretical efficiency, the dominant factor underlying the difference in runtime remains the reduced number of breakpoints explored by IPC.

As an example, Figure~\ref{fig:envelope-conductance} shows the convex envelope and the subset of breakpoints explored by IPC on the {\sc flickr$_M$} dataset. There, we can see that the breakpoints explored by IPC (the red crosses) account for only $0.33\%$ of the total number of points (the blue dots). 

Regarding the comparison of the running times, the experimental results in Table~\ref{tab:breakpoints_conductance} demonstrate the significant computational advantages of IPC over the fully parametric cut approach. IPC consistently achieves dramatic runtime improvements across all datasets, requiring only 0.13\%--16.95\% of the time needed by the fully parametric method. This efficiency stems from IPC's examining merely 0.30\%--50.00\% of the total breakpoints while still identifying the optimal minimum conductance* values. The performance gap is particularly pronounced for larger networks; for instance, on the {\sc livejournal$_M$} dataset, IPC completes in 23.07 seconds compared to the fully parametric approach's 17,384.68 seconds (approximately 5 hours). Even for smaller networks like {\sc ego-facebook$_M$}, IPC maintains its proportional advantage. We again note that, in general, there is a linear relationship between the number of explored breakpoints and the running time.

\section{Conclusions and future work}
\label{sec:conclusions}
This paper presents comprehensive computational evidence that the Incremental Parametric Cut (IPC) algorithm provides a dramatic improvement in solving the densest subgraph problem and the conductance* problem, as compared to existing methodologies. Our experimental results demonstrate that IPC significantly outperforms both heuristic methods and the fully parametric cut procedure across diverse datasets.

For the densest subgraph problem, we show that IPC is not only substantially faster than the recent heuristic approaches of Greedy++ and FISTA, but also guarantees optimal solutions. The performance gap is particularly striking for large-scale instances, where IPC processes graphs with hundreds of millions of edges in minutes, while maintaining optimality guarantees that heuristic approaches cannot provide.

A key finding of our study is the stark contrast between IPC and the fully parametric cut procedure in terms of the number of breakpoints explored. While both approaches have the same theoretical complexity, IPC examines only 2 to 14 breakpoints across all test instances, representing less than 1\% of the breakpoints explored by the fully parametric procedure. This dramatic reduction in breakpoint exploration translates to significant practical performance improvements, particularly evident in large-scale graphs where IPC terminates with an optimal solution in seconds or minutes, as compared to hours of run times for the fully parametric approach.

For the conductance* problem, our experiments demonstrate that IPC maintains its superior performance characteristics, establishing that given a subgraph guaranteed to contain the 
optimal conductance subgraph, IPC delivers this optimal solution, and very efficiently.
In particular we plan to explore applying conductance* on such potential subgraphs, resulting not only from running METIS, as done here, but also extracted from the eigenvector solution delivered by the spectral method.

The implementation of IPC provided here, along with our comprehensive test results, establishes a new benchmark for solving ratio problems on graphs. This advancement enables the practical solution of problems that were previously considered computationally intractable with exact methods.

\section*{Acknowledgments}

The research was supported in part by the AI4OPT institute NSF award 2112533.
The authors thank the Berkeley Research Computing program at the University of California, Berkeley for providing the use of the Savio computing cluster. 
The second author thanks ANID for funding through PFCHA / Doctorado Nacional / 2021-21211768.

\bibliographystyle{unsrt}
\bibliography{biblio}

\end{document}